\documentclass[10pt]{article}
\pdfoutput=1

\usepackage[utf8]{inputenc}
\usepackage[T2A]{fontenc}

\usepackage{multicol}

\input{epsf}
\usepackage{epsfig}
\usepackage{amssymb}
\usepackage{amsfonts}
\usepackage{amsbsy}
\usepackage[cmtip,all]{xy}
\usepackage{amsmath}
\usepackage{esint}
\usepackage{collectbox}
\usepackage{amscd}
\usepackage{mathrsfs}
\usepackage{amsmath,amsthm}
\usepackage{enumitem}
\usepackage{dsfont}
\usepackage{soul}
\usepackage{comment}
\usepackage{cite} 
\usepackage[most]{tcolorbox}
\usepackage{stmaryrd}

\usepackage{xcolor}
\definecolor{burgundy}{rgb}{0.5, 0.0, 0.13}
\definecolor{palette1}{rgb}{0.603922, 0.466667, 0.811765}
\definecolor{palette2}{rgb}{0.329412, 0.219608, 0.517647}
\definecolor{palette3}{rgb}{0.0156863, 0.282353, 0.333333}
\definecolor{palette4}{rgb}{0.631373, 0.211765, 0.439216}
\definecolor{palette5}{rgb}{0.92549, 0.254902, 0.462745}
\definecolor{palette6}{rgb}{1., 0.643137, 0.368627}
\definecolor{palette7}{rgb}{0.313725, 0.45098, 0.85098}
\definecolor{olive}{rgb}{0.50, 0.50, 0.0}

\usepackage[linktocpage=true,colorlinks=true,linkcolor=white!10!burgundy,citecolor=white!30!palette3,urlcolor=palette7]{hyperref}
\usepackage{accents}
\usepackage{xfrac}
\usepackage{bm}

\usepackage{tikz}
\usepackage{tikz-cd}
\usetikzlibrary{arrows}
\usetikzlibrary{arrows.meta}
\usetikzlibrary{positioning}
\usetikzlibrary{shapes}
\usetikzlibrary{fit}
\usetikzlibrary{decorations.pathmorphing,decorations.pathreplacing,decorations.markings}
\usetikzlibrary{calc}

\usepackage[font=footnotesize,labelfont=bf]{caption}

\theoremstyle{definition}

\DeclareMathAlphabet{\mathpzc}{OT1}{pzc}{m}{it}

\def\exp{{\rm exp}}

\def\I{{\rm i}}

\def\log{{\rm log}}

\def\Tr{{\rm Tr}}

\def\p{\partial}

\def\CK {{\cal K}}

\def\IC{\mathbb{C}}

\def\fg{\mathfrak{g}}

\def\fl{\mathfrak{l}}
\def\fm{\mathfrak{m}}

\def\fs{\mathfrak{s}}
\def\ft{\mathfrak{t}}

\def\fs{\mathfrak{s}}
\def\ft{\mathfrak{t}}

\def\lm{\limits}
\def\be{\begin{eqnarray}}
\def\ee{\end{eqnarray}}

\def\Al{\Delta} 

\hoffset= - 1.0in

\numberwithin{equation}{section}

\tcbset{boxrule=0.6pt,colback=white!90!palette6}

\DeclareSymbolFont{bbsymbol}{U}{bbold}{m}{n}
\DeclareMathSymbol{\bbzero}{\mathbin}{bbsymbol}{"30}
\DeclareMathSymbol{\bbone}{\mathbin}{bbsymbol}{"31}
\DeclareMathSymbol{\bbtwo}{\mathbin}{bbsymbol}{"32}
\DeclareMathSymbol{\bbthree}{\mathbin}{bbsymbol}{"33}
\DeclareMathSymbol{\bbfour}{\mathbin}{bbsymbol}{"34}
\DeclareMathSymbol{\bbfive}{\mathbin}{bbsymbol}{"35}
\DeclareMathSymbol{\bbsix}{\mathbin}{bbsymbol}{"36}
\DeclareMathSymbol{\bbseven}{\mathbin}{bbsymbol}{"37}
\DeclareMathSymbol{\bbeight}{\mathbin}{bbsymbol}{"38}
\DeclareMathSymbol{\bbnine}{\mathbin}{bbsymbol}{"39}

\def\myblue{white!40!blue}
\def\mygreen{black!40!green}

\definecolor{palette1}{rgb}{0.603922, 0.466667, 0.811765}
\definecolor{palette2}{rgb}{0.329412, 0.219608, 0.517647}
\definecolor{palette3}{rgb}{0.0156863, 0.282353, 0.333333}
\definecolor{palette4}{rgb}{0.631373, 0.211765, 0.439216}
\definecolor{palette5}{rgb}{0.92549, 0.254902, 0.462745}
\definecolor{palette6}{rgb}{1., 0.643137, 0.368627}
\definecolor{palette7}{rgb}{0.313725, 0.45098, 0.85098}

\definecolor{Xmagenta}{HTML}{D60270}
\definecolor{Xpurple}{HTML}{9B4F96}
\definecolor{Xblue}{HTML}{0038A8}

\newcommand\sqbox[1]{{
		\setbox0=\hbox{\mbox{$\Box$}}
		\setbox1=\hbox{\mbox{\raisebox{0.35ex}{\small #1}}}
		\mbox{\raisebox{-0.2ex}{\rlap{\hbox to \wd0{\hss{\box1}\hss}}\box0}}
}}
\newcommand\ssqbox[1]{{
		\setbox0=\hbox{\mbox{$\scriptstyle\Box$}}
		\setbox1=\hbox{\mbox{\raisebox{0.35ex}{\tiny #1}}}
		\mbox{\raisebox{-0.2ex}{\rlap{\hbox to \wd0{\hss{\box1}\hss}}\box0}}
}}

\begin{document}

\hfill MIPT/TH-13/26

\hfill ITEP/TH-13/26

\hfill IITP/TH-13/26

\vskip 1.5in
\begin{center}
	
    {\bf\Large Two roles of Alexander in two Kashaev phases}

	\vskip 0.2in
	\renewcommand{\thefootnote}{\fnsymbol{footnote}}
	{Dmitry Galakhov
		\footnote[2]{e-mail: d.galakhov.pion@gmail.com, galakhov@itep.ru} and  Alexei Morozov
		\footnote[3]{e-mail: morozov@itep.ru}}\\
	
	\vskip 0.2in
	\renewcommand{\thefootnote}{\roman{footnote}}
	{\small{
			\textit{
				MIPT, 141701, Dolgoprudny, Russia
			}
			\vskip 0 cm
			\textit{
				NRC “Kurchatov Institute”, 123182, Moscow, Russia
			}
			\vskip 0 cm
			\textit{
				IITP RAS, 127051, Moscow, Russia}
			\vskip 0 cm
			\textit{
				ITEP, Moscow, Russia}
	}}
\end{center}

\vskip 0.2in
\baselineskip 16pt

\centerline{ABSTRACT}

\bigskip

{\footnotesize
	The crucial feature of resurgence theory is the ambiguity of non-perturbative behavior,
	reflected either in the different choices of integration contours or in the existence of several solutions
	to Ward identities.   
	This is well illustrated by considering exactly solvable models, of which the prominent
	example is Chern-Simons theory.
	Its important chapter, which should have a direct generalization
    to arbitrary Yang-Mills, is the consideration of Wilson averages in the double-scaling 
    limit of large representation and small coupling.
    For historical reasons, we call it a Kashaev limit. 
    It possesses a natural interpretation in terms of quasiclassical/WKB approximation,
    which is, however, somewhat peculiar and thus sheds new light on the old story.
    The crucial point is the appearance of Alexander polynomials $\Al$ in two seemingly 
    opposite roles:
    the classical $A$-polynomials have common roots with $\Al$, 
    while Jones polynomials tend to $\Al^{-1}$ in the perturbative expansion.
    The consistency is provided  by the peculiar form of the {\it quantum} $A$-polynomial,
    and the resolution of the puzzle is the co-existence of two different \emph{branches} (\emph{phases}) in the quasiclassical limit --
    with non-trivial and with vanishing classical actions.
    The first leads to classical $A$-polynomials and hyperbolic volumes,
    the second -- to inverse Alexanders.

}

\bigskip

\bigskip

\tableofcontents

\bigskip

\section{Introduction}

Phase transitions could be considered one of the most picturesque and captivating phenomena in physics.
A very rough ``level 0'' explanation for why they occur in nature comes from considerations of 0d theories.
Asymptotic series mimicking perturbative expansions in parameters in QFT indicate that the actual partition function or correlators are not functions on the parameter space but rather  sections of \emph{branched} covers.
Naively, these other branches (phases) are not manifest in a perturbative expansion, yet transporting a theory along a sufficiently long loop in the parameter space might induce monodromic jumps between branches.
For low-dimensional or sufficiently simple QFTs \emph{resurgence} methods \cite{dunne_unsal_2015,aniceto_basar_schiappa_2019,dunne_2025} allow one to recover non-perturbative effects of other branches in perturbative expansions. 

An intermediate point on a path from simple towards more involved QFTs and string theory is integrable theories and TQFTs (also theories with topological subsectors).
Non-perturbative effects for those might be tamed by enhanced symmetries of the theories, so often they demonstrate integrable \cite{Duistermaat:1982vw,Blau:1990ej,Keski-Vakkuri:1991owp,Morozov:1991ds,Morozov:1991rp,Morozov:1992,Morozov:1992ea,Morozov:1994hh,Mironov:1993wi,Morozov:1995pb,Mironov:1994sf,Morozov:2005mz,Morozov:2022bnz} and even superintegrable \cite{Itoyama:2017xid,Mironov:2017och,Mironov:2022fsr,Mironov:2022lwz,Mironov:2024jdu,Mironov:2024ilq,Cassia:2025qga}\footnote{For a fresh review of ``ordinary'' superitegrability and relevant references see \cite{Turbiner:2025xjj}.} properties.
3d Chern-Simons theory is an example of a TQFT where problems of calculating expectation values of knotted and linked Wilson loop operators as well as computing partition functions on 3d manifolds are known to be solvable exactly \cite{RT1,RT2,1112.2654,Mironov:2015aia,Mironov:2015qma,Anokhina:2024lbn,Galakhov:2014sha}.
Structures of the respective asymptotic series in this case and resurgent studies of the parameter spaces were considered in \cite{gukov_marino_putrov_2016,garoufalidis_gu_marino_2023,garoufalidis_gu_marino_wheeler_2025,adams_costin_dunne_gukov_oner_2025,adams_dunne_2026}.

In this note we would like to address the following issue.
Colored Jones polynomials $J_{\CK}(r|q)$ of knots represent expectation values of knotted Wilson loops in representation $[r]$ computed via the path integral of 3d Chern-Simons theory with gauge group $SU(2)$.
Quantum parameter $q=e^{\frac{2\pi\sqrt{-1}}{k+2}}$ depends on Chern-Simons level $k$.
It is expected that Jones polynomials exhibit a quasi-classical, asymptotic behavior in a double-scaling limit, where $q=e^{\hbar}$, $\hbar\to 0$ and $r\to\infty$ so that $\log\,\ft:=r\hbar$ remains finite.
In the literature, one can find two completely different, seemingly contradictory results for this limit.
On the one hand as it was observed by R.~Kashaev \cite{Kashaev} the Jones polynomials have a natural quasi-classical, eikonal behavior reminiscent of wave functions in quantum mechanics:
\begin{equation}\label{Jassymp1}
	J_{\CK}(r|q)\longrightarrow \exp\left(S_{\CK}(\ft)/\hbar+O(\hbar^0)\right)\,,
\end{equation}
where at a specific value $\ft=1$ the exponential factor $S_{\CK}(1)$ represents the hyperbolic volume $V_{\CK}$ of the knot $\CK$ if $\CK$ is hyperbolic.
We will call this quasi-classical double-scaling limit a Kashaev limit throughout the paper.
On the other hand a famous observation by P.~Melvin and H.~Morton \cite{Melvin1995} (examined, proved and extended in numerous papers \cite{Rozansky:1994qe,BarNatan1996,1997CMaPh.183..291R,Garoufalidis:2026mvw,KLM}) indicates that if one resums the asymptotic perturbative series for $J_{\CK}(r|q)$ in $\hbar$ one arrives at a different result:
\begin{equation}\label{Jassymp2}
	J_{\CK}(r|q)\longrightarrow \frac{1}{\Al_{\CK}(\ft)}+O(\hbar)\,,
\end{equation}
where $\Delta_{\CK}$ is the Alexander polynomial of the knot $\CK$.
Moreover, expansion \eqref{Jassymp1} is expected to deliver roots of classical A-polynomials $A_{\CK}(\ell,m)$ \cite{Cooper1994Plane}:
\begin{equation}
	A_{\CK}(\ell,m)=0,\quad \ell=\frac{d}{d \ft}S_{\CK}(\ft),\quad m=\ft\,.
\end{equation}
a statement known as a direct consequence of the AJ conjecture \cite{garoufalidis_2004,le_2006,garoufalidis_le_2005} (see also \cite{garoufalidis_2010,garoufalidis_koutschan_2018,andersen_malusa_2017} for modern reviews).
The Alexander polynomial is hidden in this limit as well; for a specific value $\ell=1$ the Burde-deRham theorem \cite{de_rham_1967,Cooper1994Plane,may_2006,silver_williams_2011} implies the following factorization:
\begin{equation}
	A_{\CK}(1,\ft)=\Al_{\CK}(\ft)\times R_{\CK}(\ft)\,,
\end{equation}
where we will refer to $R_{\CK}$ as a remnant polynomial.

Despite apparent differences, two limits \eqref{Jassymp1} and \eqref{Jassymp2} ought to coexist.
The asymptotic behavior of expression \eqref{Jassymp1}, we might have called a non-perturbative Kashaev limit, follows naturally from a saddle point approximation for sums over spins that appear in a construction for Jones polynomials within the Reshetikhin-Turaev formalism \cite{RT1,RT2,1112.2654,Mironov:2015aia,Mironov:2015qma,Anokhina:2024lbn,Galakhov:2014sha}.
Also it is known that the Jones polynomial can be presented as the following series in both $r$ and $\hbar$:
\begin{equation}
	J_{\CK}(r|e^{\hbar})=\sum\lm_{k=0}^{\infty}\sum\lm_{l=k}^{\infty}a_{k,l}r^k\hbar^l=\sum\lm_{k=0}^{\infty}a_{k,k}\left(\log\,\ft\right)^k+\hbar\sum\lm_{k=0}^{\infty}a_{k,k+1}\left(\log\,\ft\right)^k+\hbar^2\sum\lm_{k=0}^{\infty}a_{k,k+2}\left(\log\,\ft\right)^k+\ldots\,.
\end{equation}
A very natural course of action in this case, which we could have called a \emph{naive} Kashaev limit, is to omit higher terms in $\hbar$ in this expansion, and then the first series sums to $\Delta_{\CK}(\ft)^{-1}$.

Here we would like to argue that the observations presented above could be accounted for by a non-homogeneous quantum A-polynomial equation for Jones polynomials:
\begin{tcolorbox}
\begin{equation}\label{main1}
	{\bf A}_{\CK}(\hat \ell,\hat m|q)J_{\CK}(r|q)={\bf R}_{\CK}(q^r|q),\quad \hat \ell\,J_{\CK}(r|q)=J_{\CK}(r+1|q),\quad \hat m\, J_{\CK}(r|q)=q^r\,J_{\CK}(r|q)\,,
\end{equation}
\end{tcolorbox}
where quantum polynomials have \emph{correct} quasi-classical limits:
\begin{equation}
	{\bf A}_{\CK}(\ell,m|1)=A_{\CK}(\ell,m),\quad {\bf R}_{\CK}(\ft|1)=R_{\CK}(\ft)=\frac{A_{\CK}(1,\ft)}{\Al_{\CK}(\ft)}\,.
\end{equation}

Also statement \eqref{main1} could be reformulated in the following way:\footnote{
In fact, one could derive the whole two-parametric family of \emph{equivalent} equations:
\begin{equation}
	\left(A_{\CK}\left(e^{\ft S'(\ft)},\ft\right)-e^{a\ft S'(\ft)}A_{\CK}\left(1,\ft\right)\right)J_{\CK}+e^{a\ft S'(\ft)}R(\ft)\left(\Al_{\CK}(\ft)J_{\CK}-e^{b \ft S'(\ft)}\right)=O(\hbar)\,.
\end{equation}
}
\begin{tcolorbox}
\begin{equation}\label{main2}
	\begin{array}{c}
		\begin{tikzpicture}
			\node {$\left(A_{\CK}\left(e^{\ft S'(\ft)},\ft\right)-A_{\CK}\left(1,\ft\right)\right)J_{\CK}+R(\ft)\left(\Al_{\CK}(\ft)J_{\CK}-{\bf\color{burgundy} 1}\right)=O(\hbar)\,.$};
			\foreach \x/\y/\h in {-4.8/2.6/0.3} {
				\draw[ultra thin, fill] (\x,\h) to[out=90,in=180] (\x+0.3,\h+0.17) -- (0.5*\x+0.5*\y-0.3,\h+0.17) to[out=0,in=270] (0.5*\x+0.5*\y,\h+0.3) to[out=270,in=0] (0.5*\x+0.5*\y-0.3,\h+0.13) -- (\x+0.3,\h+0.13) to[out=180,in=90] (\x,\h);
				\draw[ultra thin, fill] (\y,\h) to[out=90,in=0] (\y-0.3,\h+0.17) -- (0.5*\x+0.5*\y+0.3,\h+0.17) to[out=180,in=270] (0.5*\x+0.5*\y,\h+0.3) to[out=270,in=180] (0.5*\x+0.5*\y+0.3,\h+0.13) -- (\y-0.3,\h+0.13) to[out=0,in=90] (\y,\h);
				\node[above] at (0.5*\x+0.5*\y,\h+0.3) {$\scriptstyle \left(A_{\CK}\left(e^{\ft S'(\ft)},\ft\right)-A_{\CK}\left(1,\ft\right)+R(\ft)\Al_{\CK}(\ft)\right)J_{\CK}=0$};
			}
			\foreach \x/\y/\h in {1.1/3.4/-0.2} {
				\draw[ultra thin, fill] (\x,\h) to[out=270,in=180] (\x+0.3,\h-0.17) -- (0.5*\x+0.5*\y-0.3,\h-0.17) to[out=0,in=90] (0.5*\x+0.5*\y,\h-0.3) to[out=90,in=0] (0.5*\x+0.5*\y-0.3,\h-0.13) -- (\x+0.3,\h-0.13) to[out=180,in=270] (\x,\h);
				\draw[ultra thin, fill] (\y,\h) to[out=270,in=0] (\y-0.3,\h-0.17) -- (0.5*\x+0.5*\y+0.3,\h-0.17) to[out=180,in=90] (0.5*\x+0.5*\y,\h-0.3) to[out=90,in=180] (0.5*\x+0.5*\y+0.3,\h-0.13) -- (\y-0.3,\h-0.13) to[out=0,in=270] (\y,\h);
				\node[below] at (0.5*\x+0.5*\y,\h-0.3) {$\scriptstyle \Al_{\CK}(\ft)J_{\CK}-1=0$};
			}
			\begin{scope}[shift={(-1.35,0.95)}]
				\draw[thick, burgundy] (-0.5,-0.15) -- (0.5,0.15);
			\end{scope}
			\begin{scope}[shift={(0.05,0.95)}]
				\draw[thick, burgundy] (-0.5,-0.15) -- (0.5,0.15);
			\end{scope}
		\end{tikzpicture}
	\end{array}
\end{equation}
\end{tcolorbox}
In this formulation, two roles of the Alexander polynomial for two asymptotic branches become more transparent.
When the classical action is zero (this is governed by the \emph{Abelian} branch/phase), also $S'(\ft)=0$, and the whole first bracket vanishes.
The equation reduces to a simple part emphasized in the bottom right corner and produces branch \eqref{Jassymp2}.
On the other hand, for the \emph{non-Abelian} branches/phases, the very Jones polynomial becomes exponentially large and suppresses the unity term in the second bracket, as it is emphasized in the top left corner.
In this case, the Alexander polynomial is promoted to the A-polynomial, whose roots in $\ell$ define the eikonal behavior \eqref{Jassymp1} of this branch of the Jones polynomial.

The paper is organized as follows.
We start Sec.~\ref{sec:toy} by recalling some simple examples of ordinary series reflecting asymptotic properties similar to those of the Jones polynomials and test on them the concepts proposed in the paper.
In Sec.~\ref{sec:examples} we consider explicit examples of the trefoil knot and of the figure-eight knot.
In Sec.~\ref{sec:equations} we review constructions of difference equations for the Jones polynomials leading to relations \eqref{main1} and \eqref{main2}.
Finally, in Sec.~\ref{sec:why} we propose some arguments clarifying why it is precisely the Alexander polynomial appears in these two roles in these two seemingly different branches/phases.


\section{Toy model} \label{sec:toy}

We would like to begin our discussion with a toy model of a simple asymptotic series exhibiting similar properties to those of the knot polynomials we discuss in what follows.

As a warm-up exercise, consider the following series:
\begin{equation}
	f^{(1)}(r,\hbar):=1+\frac{\hbar r}{1!}+\frac{\hbar^2 r(r-1)}{2!}+\frac{\hbar^3 r(r-1)(r-2)}{3!}+\ldots=\sum_{k=0}^r \frac{r!}{k! (r-k)!} \hbar^k = (1+\hbar)^r\,.
\end{equation}
The naive KL following the series definition reads:
\begin{equation}
\tilde f^{(1)} = \sum\lm_{k=0}^{\infty}\frac{\hbar^k r^k}{k!}= \sum\lm_{k=0}^{\infty}\frac{\left(\log\,\ft\right)^k}{k!}=e^{\log \ft} =\ft\,.
\label{KL1}
\end{equation}
In this case it coincides with the limit from the analytically continued function:
\begin{equation}
	(1+\hbar)^n=\exp\left(\frac{\log\,\ft}{\hbar}\log(1+\hbar)\right)=e^{\log\,\ft+O(\hbar)}=\ft+O(\hbar)\,.
\end{equation}

Now let us turn to a more singular case of the series:
\begin{equation}
	f^{(2)}(r,\hbar):=1+\hbar r+\hbar^2 r(r-1)+\hbar^3 r(r-1)(r-2)+\ldots=\sum_{k=0}^\infty \frac{r!}{(r-k)!} \hbar^k\,.
	\label{f2}
\end{equation}
We should stress that, despite the fact that we have deleted suppressing factorials from the denominators, for each finite positive integer $r$ this series is terminated by a zero contribution in the numerator and gives a finite polynomial.

It is straightforward to derive the naive Kashaev limit:
\begin{equation}
	\tilde f^{(2)}=1+\hbar r+\hbar^2 r^2+\hbar^3 r^3+\ldots=\sum_{k=0}^{\infty} \left( \hbar r\right)^k=\frac{1}{1-\hbar r}=\frac{1}{1-\log \, \ft}\,.
	\label{KL2}
\end{equation}

To approach a calculation of its asymptotic behavior from another angle, we would like to represent the series in integral form.
For this purpose, it turns out, the standard Borel presentation is suitable, since the Borel pre-image of series $f^{(2)}$ is the series $f^{(1)}$:
\begin{equation}
	\sum_{k=0}^\infty \frac{r!}{(r-k)!} \hbar^k\to \sum_{k=0}^\infty \frac{k!r!}{k!(r-k)!} \hbar^k\to \int\lm_0^{+\infty}ds\;e^{-s} \sum_{k=0}^\infty \frac{r!}{k!(r-k)!} \left(s\hbar\right)^k=\int\lm_0^{\infty}ds\; e^{-s}\left(1+\hbar s\right)^{r}=:f^{(2)}_B\,.
\end{equation}
Now we would like to study the behavior of this integral representation in the limit $\hbar\to 0$ with $r=\xi/\hbar$.
By redefining the variable as $z:=s/\hbar$, we arrive at a 0d QFT representation of this integral:
\begin{equation}\label{fB2}
	f_B^{(2)}(z,\hbar)=\frac{1}{\hbar}\int\lm_0^{+\infty}dz\;\exp\left(-\frac{S(z,\xi)}{\hbar}\right),\quad S(z,\xi):=\left(z-\xi\,\log(1+z)\right)\,.
\end{equation}

We can determine a saddle point $z_*$ by solving ``equations of motion'':
\begin{equation}
	\p_zS(z,\ft)=1-\frac{\xi}{1+z_*}=0\;\Rightarrow\; z_*=\xi-1\,.
\end{equation}

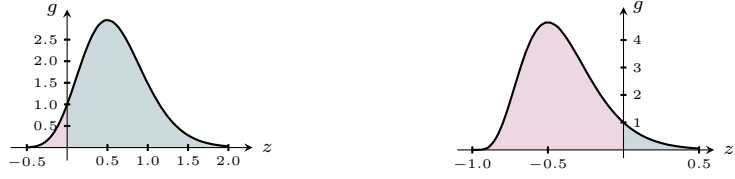
\begin{figure}
	\centering
	\begin{tikzpicture}
		\node at (0,0) {$\begin{array}{c}
				\begin{tikzpicture}[yscale=0.571429, xscale=1.06667]
					\draw[draw=none, fill=white!80!palette4] (-0.5,0) -- (-0.5,0.00453) to[out=5.39971,in=197.172] (-0.42,0.01886) to[out=17.1721,in=217.524] (-0.34,0.05885) to[out=37.5242,in=236.998] (-0.26,0.14711) to[out=56.9983,in=243.864] (-0.22,0.2172) to[out=63.8637,in=248.906] (-0.18,0.30827) to[out=68.9056,in=252.537] (-0.14,0.42217) to[out=72.5372,in=256.136] (-0.08,0.63717) to[out=76.1365,in=257.683] (-0.04,0.8087) to[out=77.6827,in=258.634] (0.,1.) -- (0,0) -- cycle;
					\draw[draw=none, fill=white!80!palette3] (0,0) -- (0.,1.) to[out=78.8961,in=259.302] (0.18,1.97925) to[out=79.3019,in=257.162] (0.26,2.37899) to[out=77.1618,in=251.704] (0.34,2.69135) to[out=71.704,in=234.848] (0.42,2.88595) to[out=54.8484,in=168.975] (0.5,2.95051) to[out=-11.0254,in=121.585] (0.58,2.89034) to[out=-58.4154,in=110.016] (0.66,2.72444) to[out=-69.9838,in=105.978] (0.74,2.48004) to[out=-74.0224,in=104.476] (0.82,2.18704) to[out=-75.524,in=105.92] (1.04,1.34212) to[out=-74.0797,in=107.966] (1.12,1.07384) to[out=-72.0344,in=110.913] (1.2,0.84102) to[out=-69.0866,in=114.918] (1.28,0.64573) to[out=-65.0816,in=120.135] (1.36,0.48672) to[out=-59.8648,in=126.624] (1.44,0.36058) to[out=-53.3762,in=134.216] (1.52,0.26286) to[out=-45.7839,in=150.503] (1.68,0.13362) to[out=-29.4974,in=163.777] (1.84,0.06438) to[out=-16.2229,in=171.191] (2.,0.02958) -- (2,0) -- cycle;
					\draw[thick] (-0.5,0.00453) to[out=5.39971,in=197.172] (-0.42,0.01886) to[out=17.1721,in=217.524] (-0.34,0.05885) to[out=37.5242,in=236.998] (-0.26,0.14711) to[out=56.9983,in=243.864] (-0.22,0.2172) to[out=63.8637,in=248.906] (-0.18,0.30827) to[out=68.9056,in=252.537] (-0.14,0.42217) to[out=72.5372,in=256.136] (-0.08,0.63717) to[out=76.1365,in=257.683] (-0.04,0.8087) to[out=77.6827,in=258.634] (0.,1.);
					\draw[thick] (0.,1.) to[out=78.8961,in=259.302] (0.18,1.97925) to[out=79.3019,in=257.162] (0.26,2.37899) to[out=77.1618,in=251.704] (0.34,2.69135) to[out=71.704,in=234.848] (0.42,2.88595) to[out=54.8484,in=168.975] (0.5,2.95051) to[out=-11.0254,in=121.585] (0.58,2.89034) to[out=-58.4154,in=110.016] (0.66,2.72444) to[out=-69.9838,in=105.978] (0.74,2.48004) to[out=-74.0224,in=104.476] (0.82,2.18704) to[out=-75.524,in=105.92] (1.04,1.34212) to[out=-74.0797,in=107.966] (1.12,1.07384) to[out=-72.0344,in=110.913] (1.2,0.84102) to[out=-69.0866,in=114.918] (1.28,0.64573) to[out=-65.0816,in=120.135] (1.36,0.48672) to[out=-59.8648,in=126.624] (1.44,0.36058) to[out=-53.3762,in=134.216] (1.52,0.26286) to[out=-45.7839,in=150.503] (1.68,0.13362) to[out=-29.4974,in=163.777] (1.84,0.06438) to[out=-16.2229,in=171.191] (2.,0.02958);
					\draw[-stealth] (-0.7,0) -- (2.3,0);
					\draw[-stealth] (0,-0.3) -- (0,3.2);
					\node[left] at (0,3.2) {$\scriptstyle g$};
					\node[right] at (2.3,0) {$\scriptstyle z$};
					\foreach \x in {-0.5,0.5,1.0,1.5,2.0} {
						\draw[thick] (\x,-0.08) -- (\x,0.08);
						\node[below] at (\x,0) {\tiny $\x$};
					}
					\foreach \y in {0.5,1.0,1.5,2.0,2.5} {
						\draw[thick] (-0.04,\y) -- (0.04,\y);
						\node[left] at (0,\y) {\tiny $\y$};
					}
				\end{tikzpicture}
			\end{array}$};
		\node at (6,0) {$\begin{array}{c}
				\begin{tikzpicture}[yscale=0.363636, xscale=2.]
					\draw[draw=none, fill=white!80!palette4] (-1,0) -- (-1.,0.) to[out=0.00002,in=234.45] (-0.926,0.02332) to[out=54.4498,in=259.796] (-0.878,0.17575) to[out=79.7962,in=263.538] (-0.85,0.37322) to[out=83.5378,in=265.448] (-0.818,0.71267) to[out=85.4484,in=266.338] (-0.786,1.16312) to[out=86.3379,in=266.738] (-0.754,1.69534) to[out=86.7378,in=266.72] (-0.692,2.80589) to[out=86.7205,in=265.598] (-0.628,3.80263) to[out=85.5985,in=264.179] (-0.596,4.17159) to[out=84.1786,in=261.123] (-0.564,4.43459) to[out=81.1227,in=251.764] (-0.532,4.58855) to[out=71.7637,in=174.715] (-0.5,4.63791) to[out=-5.28538,in=109.388] (-0.468,4.59258) to[out=-70.6124,in=98.2085] (-0.404,4.27347) to[out=-81.7915,in=96.259] (-0.34,3.7525) to[out=-83.741,in=96.5051] (-0.182,2.26039) to[out=-83.4949,in=97.5928] (-0.118,1.73704) to[out=-82.4072,in=98.4663] (-0.082,1.48025) to[out=-81.5338,in=101.266] (0.,1.) -- (0,0) -- cycle;
					\draw[draw=none, fill=white!80!palette3] (0,0) -- (0.,1.) to[out=-78.5792,in=104.607] (0.06,0.73443) to[out=-75.3931,in=107.385] (0.1,0.59247) to[out=-72.615,in=110.797] (0.14,0.4748) to[out=-69.2029,in=114.929] (0.18,0.37816) to[out=-65.0708,in=119.83] (0.22,0.29947) to[out=-60.1699,in=125.474] (0.26,0.23588) to[out=-54.5263,in=138.305] (0.34,0.14419) to[out=-41.6948,in=151.081] (0.42,0.08658) to[out=-28.9187,in=160.886] (0.5,0.05117) -- (0.5,0) -- cycle;
					\draw[thick] (-1.,0.) to[out=0.00002,in=234.45] (-0.926,0.02332) to[out=54.4498,in=259.796] (-0.878,0.17575) to[out=79.7962,in=263.538] (-0.85,0.37322) to[out=83.5378,in=265.448] (-0.818,0.71267) to[out=85.4484,in=266.338] (-0.786,1.16312) to[out=86.3379,in=266.738] (-0.754,1.69534) to[out=86.7378,in=266.72] (-0.692,2.80589) to[out=86.7205,in=265.598] (-0.628,3.80263) to[out=85.5985,in=264.179] (-0.596,4.17159) to[out=84.1786,in=261.123] (-0.564,4.43459) to[out=81.1227,in=251.764] (-0.532,4.58855) to[out=71.7637,in=174.715] (-0.5,4.63791) to[out=-5.28538,in=109.388] (-0.468,4.59258) to[out=-70.6124,in=98.2085] (-0.404,4.27347) to[out=-81.7915,in=96.259] (-0.34,3.7525) to[out=-83.741,in=96.5051] (-0.182,2.26039) to[out=-83.4949,in=97.5928] (-0.118,1.73704) to[out=-82.4072,in=98.4663] (-0.082,1.48025) to[out=-81.5338,in=101.266] (0.,1.);
					\draw[thick] (0.,1.) to[out=-78.5792,in=104.607] (0.06,0.73443) to[out=-75.3931,in=107.385] (0.1,0.59247) to[out=-72.615,in=110.797] (0.14,0.4748) to[out=-69.2029,in=114.929] (0.18,0.37816) to[out=-65.0708,in=119.83] (0.22,0.29947) to[out=-60.1699,in=125.474] (0.26,0.23588) to[out=-54.5263,in=138.305] (0.34,0.14419) to[out=-41.6948,in=151.081] (0.42,0.08658) to[out=-28.9187,in=160.886] (0.5,0.05117);
					\draw[-stealth] (-1.1,0) -- (0.6,0);
					\draw[-stealth] (0,-0.3) -- (0,5.2);
					\node[right] at (0,5.2) {$\scriptstyle g$};
					\node[right] at (0.6,0) {$\scriptstyle z$};
					\foreach \x in {-1.0,-0.5,0.5} {
						\draw[thick] (\x,-0.13) -- (\x,0.13);
						\node[below] at (\x,0) {\tiny $\x$};
					}
					\foreach \y in {1,2,3,4} {
						\draw[thick] (-0.02,\y) -- (0.02,\y);
						\node[right] at (0,\y) {\tiny $\y$};
					}
				\end{tikzpicture}
			\end{array}$};
	\end{tikzpicture}
	\caption{a) exemplary plot of function $g(z)=e^{-\frac{z}{0.1}} (1+z)^{\frac{\bf\color{burgundy}1.5}{0.1}}$; b) exemplary plot of function $g(z)=e^{-\frac{z}{0.1}} (1+z)^{\frac{\bf\color{burgundy} 0.5}{0.1}}$.}\label{fig:plots}
\end{figure}

The standard prescription \cite{Howls1992} of estimating the asymptotics of integral \eqref{fB2} is to expand the integration cycle over Lefschetz thimbles associated to saddle points {\bf and} the steepest descent paths to and from the endpoints if they are not located at singularities.
To simplify our consideration and avoid constructing steepest descent paths in the complex plane,  we assume all the variables are real.
In this case, the paths all lie on the real line.
There are two possible regimes (``phases''):
\begin{enumerate}
	\item If $\xi>1$ then $z_*\in[0,+\infty)$.
	In this case, the Lefschetz thimble covers the whole integration region $[-1,+\infty)$, and the steepest descent path from the endpoint corresponds to the shaded region to the left in Fig.~\ref{fig:plots}(a). 
	Strictly speaking, an expansion of the integration region $[0,+\infty)$ suggests that one should subtract from the saddle point contribution a contribution from the endpoint.
	However, the saddle point contribution clearly dominates the integral asymptotic behavior in this case:
	\begin{equation}\label{asympt1}
		f_B^{(2)}\underset{\hbar\to 0}{\sim}\frac{1}{\hbar}\int\lm_{-\infty}^{+\infty}dz\;\exp\left(\frac{\xi\,\log\xi+1-\xi}{\hbar}-\frac{z^2}{2\xi\hbar}\right)=\sqrt{\frac{2\pi\xi}{\hbar}}e^{\frac{\xi\,\log\xi+1-\xi}{\hbar}}\,.
	\end{equation}
	\item If $\xi<1$ then $z_*\not\in[0,+\infty)$.
	In this case, the integration cycle $[0,+\infty)$ coincides with the steepest descent path from the endpoint  (see Fig.~\ref{fig:plots}(b)).
	Therefore we approximate the integral by its perturbative expansion around $z=0$:
	\begin{equation}\label{asympt2}
		f_B^{(2)}\underset{\hbar\to 0}{\sim}\frac{1}{\hbar}\int\lm_{0}^{\infty}e^{-\frac{1-\xi}{\hbar}z+O(z^2)}=\frac{1}{1-\xi}\,.
	\end{equation}
\end{enumerate}
We see clearly that in the second regime when the saddle point does not contribute, the asymptotic result coincides with the naive Kashaev limit \eqref{KL2}.

An integral representation of a series is a canonical way to extend it analytically beyond the convergence radius or even to deal with asymptotic series.
Another ``competing'' approach is to use differential/difference equations to parallel transport a function to a desired region.
Series $f^{(2)}$ clearly satisfies the following difference equation:
\begin{equation}\label{toy}
	f^{(2)}_\hbar(r+1)=1+\hbar(r+1)f^{(2)}_\hbar(r)\,.
\end{equation}
To observe how both asymptotic solutions \eqref{asympt1} and \eqref{asympt2} emerge, let us substitute here directly $r=\xi/\hbar$:
\begin{equation}\label{f_2_eq}
	f^{(2)}_\hbar(\xi+\hbar)=1+(\xi+\hbar)f^{(2)}_\hbar(\xi)\,.
\end{equation}
Usually a difference equation has infinitely many solutions.
It suffices to substitute any parameter in a given parametric solution by a periodic function to obtain a new solution.
Yet let us search for a solution in an eikonal form:
\begin{equation}
	f_\hbar^{(2)}(\xi)=e^{\frac{S(\xi)}{\hbar}}\left(\Delta(\xi)^{-1}+\hbar \nu_{\hbar}(\xi)\right)\,,
\end{equation}
where $\nu$ is a series in $\hbar$.
Substituting this ansatz into \eqref{f_2_eq}, we obtain the following expansion:
\begin{equation}
	\left(\left(e^{S'(\xi)}-\xi\right)\Delta(\xi)^{-1}-e^{-\frac{S(\xi)}{\hbar}}\right)+\hbar\left(\left(e^{S'(\xi)-\xi}\right)\nu_0(\xi)-\frac{e^{S'(\xi)}}{2}\left(-\frac{2 \Delta '(\xi )}{\Delta(\xi )^2}+\Delta (\xi )^{-1} S''(\xi )\right)-\Delta (\xi )^{-1}\right)+O(\hbar^2)=0\,.
\end{equation}
Clearly, this equation has two regimes as well:
\begin{enumerate}
	\item An eikonal regime, when $S\neq 0$, then the exponentially suppressed term in the first bracket could be omitted $e^{-\frac{S(\xi)}{\hbar}}\to 0$.
	The zeroth and the first orders deliver asymptotics  of \eqref{asympt1} with
	\begin{equation}
		S(\xi)=\xi\;\log\,\xi-\xi+C_1,\quad \Delta(\xi)=C_2/\sqrt{\xi}\,.
	\end{equation}
	The full solution is
	\begin{equation}
	f^{(2)}(\xi) = const\cdot \exp\left(\frac{\xi\;\log\,\xi-\xi}{\hbar}\right)\sqrt{\xi}
	\cdot\left(1 + \frac{\hbar}{12\xi}+\frac{\hbar^2}{288 \xi^2}-\frac{139 \hbar^3}{2^4  3^3 5!\,\xi^3}
	-\frac{571\hbar^4}{2^7 3^3 6!\, \xi^4}+\frac{29\cdot 5651 \hbar^5}{2^9  3^4  7!\,\xi^5}
	+ \ldots\right)
	\end{equation}
	\item A perturbative regime, when $S=0$.
	In this case term $e^{-\frac{S(\xi)}{\hbar}}=1$ is not suppressing, and the zeroth order approximation delivers the naive Kashaev limit:
	\begin{equation}\label{naiveKLtoy}
		S(\xi)=0,\quad \Delta(\xi)=1-\xi\,.
	\end{equation}
	The full solution in this case is
	\begin{equation}
	f^{(2)}(\xi)=\frac{1}{1- \xi}-\frac{\hbar\xi}{(1-\xi)^3}+\frac{\hbar^2 \xi(\xi+2)}{(1-\xi)^5}
	-\frac{\hbar^3\xi (\xi^2+8 \xi+6)}{(1- \xi)^7}
	+\frac{\hbar^4\xi(\xi^3+22 \xi^2+58 \xi+24)}{(1- \xi)^9}-\ldots\,.
	\end{equation}
	These terribly-looking functions are just the elementary corrections, obvious from (\ref{f2}):
	\begin{equation}
	\begin{aligned}
	&f^{(2)}(r,\hbar):=1+\hbar r+\hbar^2 r(r-1)+\hbar^3 r(r-1)(r-2)+\ldots =\\
	&= \sum_{k=0}  (r\hbar)^k - \hbar \sum_{k=1} \left(\sum_{j=0}^{k} j\right) (r\hbar)^{k}
	+ \hbar^2\sum_{k=2}\left(\sum_{0\leq j< j'\leq k}jj'\right)  (r\hbar)^{k-1}-\\
	&- \hbar^3 \sum_{k=3}\left(\sum_{0\leq j< j'<j''\leq k}jj'j''\right)  (r\hbar)^{k-2} + \ldots =\\
	&= \sum_k \xi^k - \hbar \sum_k \frac{k(k+1)}{2}\xi^k + \ldots
	= \frac{1}{1-\xi} -\frac{\hbar\xi}{(1-\xi)^3}  +\frac{\hbar^2 \xi(\xi+2)}{(1-\xi)^5}
	-\frac{\hbar^3\xi (\xi^2+8 \xi+6)}{(1- \xi)^7} + \ldots\,.
	\end{aligned}
	\end{equation}

\end{enumerate}

To conclude this section, let us note that our toy difference equation \eqref{toy} could also be brought to the form of \eqref{main1}:
\begin{equation}
	\begin{aligned}
		& {\bf A}(\hat \ell,\hat \xi|\hbar)f_\hbar^{(2)}={\bf R}(r|\hbar),\; \hat\ell\, f_\hbar^{(2)}(r)=f_\hbar^{(2)}(r+1),\; \hat\xi\, f_\hbar^{(2)}(r)=\hbar r\,f_\hbar^{(2)}(r),\; {\bf A}(\hat \ell,\hat \xi|\hbar)=\hat\ell-\hat\xi-\hbar,\; {\bf R}(r|\hbar)=1 \,.
	\end{aligned}
\end{equation}
The classical ``A-polynomial'' and the ``remnant polynomial'' read in this case $A(\ell,\xi)=\ell-\xi$, $R(\xi)=1$.
Then we find for the ``Alexander polynomial'' (cf. \eqref{naiveKLtoy}):
\begin{equation}
	\Delta(\xi)=\frac{A(1,\xi)}{R(\xi)}=1-\xi\,.
\end{equation}


\section{Simple knot examples}\label{sec:examples}

\subsection{Trefoil knot}

The Jones polynomial for the trefoil knot $3_1$ is given by the following expression \cite{garoufalidis_2004}:
\begin{equation}\label{J_3_1}
	J_{3_1}(r|q)=\sum\lm_{k=0}^{\infty}(-1)^kq^{k(k+3)}\prod\lm_{s=1}^k\{q^{r+s}\}\{q^{r-s}\}\,,
\end{equation}
where $\{x\}:=x-x^{-1}$.

First, let us address the construction of the naive Kashaev limit.
It could be done in a somewhat simple and naive way in this case as one fixes $q^r=\ft$ and sends all the remaining $r$-independent powers of $q$ to 1.
The result reads:
\begin{equation}\label{J_3_1_nKL}
	\tilde J_{3_1}(r|q)=\sum\lm_{k=0}^{\infty}(-1)^k\{\ft\}^{2k}=\frac{1}{1+\{\ft\}^2}=\frac{1}{\ft^2-1+\ft^{-2}}=\frac{1}{\Al_{3_1}(\ft)}\,,
\end{equation}
where $\Al_{3_1}(\ft)$ is the Alexander polynomial of $3_1$.

A less naive approach is to estimate the sum \eqref{J_3_1} by summing around a saddle term.
To do so we rewrite expression \eqref{J_3_1} in a form involving q-Pochhammer symbols explicitly:
\begin{equation}\label{J_3_1_q_Poch}
	J_{3_1}(r|q)=\sum\lm_{k=0}^{\infty}(-1)^kq^{k(k+3)-2rk}\frac{(q^{2r+2}|q^2)_{\infty} (q^{2r-2k}|q^2)_{\infty}}{(q^{2r+2k+2}|q^2)_{\infty}(q^{2r}|q^2)_{\infty}}\,,
\end{equation}
where
\begin{equation}
	(x|q)_{\infty}:=\prod\lm_{n=0}^{\infty}(1-x q^n)\,.
\end{equation}
In the limit $q=e^{\hbar}$ with $\hbar\to 0$ the q-Pochhammer symbol has the following asymptotic:
\begin{equation}
	(x|e^{\hbar})_{\infty}=\exp\left(\frac{1}{\hbar}{\rm Li}_2(x)-\frac{1}{2}\log(1-x)+\frac{x}{12(1-x)}\hbar+O(\hbar^2)\right)\,.
\end{equation}
Loosely speaking, in this quasi-classical limit $q^2=e^{\hbar}$ with $\hbar\to 0$, expression \eqref{J_3_1_q_Poch} could be rewritten as an effective 0d path integral over the effective field $\log\,z:=\hbar k$, $2m=\hbar r$:
\begin{equation}\label{J_3_1_int}
	J_{3_1}\sim\int\lm_0^{+\infty} dz\,e^{\frac{S(z,m)}{\hbar}},\quad S(z,m)=\pi\I\, \log\,z+\frac{1}{2}\left(\log\,z\right)^2-\log\,m^2\;\log\,z+{\rm Li}_2\left(\frac{m^2}{z}\right)-{\rm Li}_2\left(m^2z\right)\,.
\end{equation}
A saddle point for this action is defined by the following equation:
\begin{equation}
	e^{z\frac{d}{dz}S(z,m)}-1=-\frac{z_* \left(t^4-t^2 z_*+1\right)}{t^2}=0\,.
\end{equation}
This equation has two apparent roots; however $z_*=0$ is spurious, as it delivers a divergent action.

An expectation value for the shift operator is 
\begin{equation}
	\ell=e^{\frac{m}{2}\frac{d}{dm}S(z_*,m)}=\frac{t^2 z_*-1}{t^2-z_*}=-m^6\,.
\end{equation}
So for the classical A-polynomial in this case we obtain the following expression:
\begin{equation}
	A_{3_1}^{(0)}(\ell,m)=\ell+m^6\,.
\end{equation}
Knot $3_1$ is not hyperbolic, so the extremal action $S(z_*,m)$ does not represent the hyperbolic volume for this knot.

The Jones polynomial satisfies the following difference equation in this case:
\begin{equation}
	\left(1-q^{2 r}\right)J_{3_1}(r|q)+ q^{6 r-2}\left(1-q^{2 r-2}\right)J_{3_1}(r-1|q)=q^{2 r-2}\left(1-q^{4 r-2}\right) \,,
\end{equation}
that could be rewritten in the canonical form \eqref{main1}:
\begin{equation}
	{\bf A}_{3_1}(\hat\ell,\hat m|q)=(1-\hat m^2)+q^{-2}\hat m^6\left(1-q^{-2}\hat m^2\right)\hat\ell^{-1},\quad {\bf R}_{3_1}=q^{2 r-2}\left(1-q^{4 r-2}\right)\,.
\end{equation}
In the quasi-classical limit these polynomials become:
\begin{equation}
	A_{3_1}(\ell,m)=(1-m^2)(1+m^6\ell^{-1})=(1-m^2)\ell^{-1}A_{3_1}^{(0)}(\ell,m),\quad R_{3_1}(m)=m^2(1-m^4)\,,
\end{equation}
where we should note that the classical A-polynomial coincides with the one obtained via the saddle point approximation up to an inessential factor.
It is easy to observe that the classical A-polynomial factorizes, with the Alexander polynomial as one of the factors:
\begin{equation}
	A_{3_1}(1,\ft)=(1-\ft^2)(1+\ft^6)=(1-\ft^2)(1+\ft^2)(1-\ft^2+\ft^4)=R_{3_1}(\ft)\times\underbrace{\left(\ft^2-1+\ft^{-2}\right)}_{\Al_{3_1}(\ft)}\,.
\end{equation}

To conclude this subsection, let us stress that the perturbative expansion \eqref{J_3_1_nKL} could be derived in a more rigorous way by estimating the steepest descent contribution from the lower summation limit $k=0$ in \eqref{J_3_1}.
Using similarities between sum \eqref{J_3_1} and integral \eqref{J_3_1_int} we expand the action near the point $z=1$:
\begin{equation}
	S(z,\ft)=\log(-\{\ft\}^2)\;(z-1)+O\left((z-1)^2\right)\,.
\end{equation}
Yet, applying to this expression a summation over $k$ with $z=1+\hbar k+O(\hbar^2)$, we arrive at the sum \eqref{J_3_1_nKL}.


\subsection{Figure-eight knot}

We deal with the case of the figure-eight knot $4_1$ in the same fashion.
The Jones polynomial of $4_1$ reads:
\begin{equation}\label{J41}
	J_{4_1}(r|q)=\sum_{s=0}^r \frac{\prod_{j=-s}^s \{q^{r+j+1}\}}{\{q^{r+1}\}}\,.
\end{equation}
This formula follows from the differential expansion for HOMFLY polynomials \cite{Dunfield:2006,Itoyama:2012,Zhu:2013,Morozov:2012JHEP,Morozov:2013JETPL,Mironov:2013AIP,Gorsky:2013,Arthamonov:2014TMP,Arthamonov:2014JHEP,Anokhina:2014NPB,Anokhina:2014MPLA,Mironov:2013TMP,Mironov:2013EPJC,Mironov:2014NPB},
which for $\CK=4_1$ and symmetric representations is especially simple
(the defect \cite{Konodef} is zero and all the $F$-factors are  unity):
\begin{equation}\label{DE41}
	\begin{aligned}
		&{\cal H}_{[r]}^{4_1}(q,A) =
		\sum_{s=0}^r \frac{[r]!}{[s]![r-s]!}\prod_{j=0}^{s-1}\{Aq^{r+j}\}\{Aq^{j-1}\} \
		\stackrel{A=q^2}{\Longrightarrow} \\
		&J^{4_1}(r|q) := {\cal H}_{[r]}^{4_1}(q,A=q^2) =
		\sum_{s=0}^r     \prod_{j=1}^s \frac{\{q^{r-s+j+1}\}}{\{q^j\}} \prod_{j=0}^{s-1}\{q^{r+j+2}\}\{q^{j+1}\}
		= \sum_{s=0}^r \prod_{\stackrel{j\neq 0}{j=-s}}^s \{q^{r+j+1}\}\,.
	\end{aligned}
\end{equation}

To calculate the naive Kashaev limit in this case, we again substitute $q^r=\ft$ and eliminate all the other powers by setting $q^{\#}=1$.
The series \eqref{J41} converges to the inverse Alexander polynomial of knot $4_1$:
\begin{equation}\label{J41_naive}
	\tilde J^{4_1} = \sum_{s=0}^\infty \{\mathfrak{t}\}^{2s}
= \frac{1}{1-\{\mathfrak{t}\}^2} = \frac{1}{3-\mathfrak{t}^2 - \mathfrak{t}^{-2}}
= \frac{1}{{\Al}_{4_1}(\mathfrak{t})}\,.
\end{equation}

To estimate an asymptotic behavior of series \eqref{J41}, we again re-express products of quantum numbers as quantum dilogarithms:
\begin{equation}
	J_{4_1}(r|q) = \frac{1}{\{q^{r+1}\}} \sum_{s=0}^\infty \prod_{j=-s}^s \{q^{r+j+1}\}
	= \frac{1}{\{q^{r+1}\}} \sum_{s=0}^\infty q^{(r+1)(2s+1)}
	\frac{(q^{-2r-2s-2}|q^2)_\infty}{(q^{-2r+2s}|q^2)_\infty}\,.
\end{equation}
Replacing the sum over integers by an integral in the limit $q^2=e^{\hbar}$ with $\hbar\to 0$, $q^r=m$, $q^{2s}=z$, one estimates this series as a partition function in a 0d theory:
\begin{equation}\label{S41}
J_{4_1} \sim\int\lm_0^{+\infty}  \exp\left(\frac{S(z,m)}{\hbar}\right)\,dz,\quad
S(z,m) = \log \,m^2  \cdot \log z
+ {\rm Li}_2\left(\frac{1}{z m^2}\right)
- {\rm Li}_2\left(\frac{z}{m^2}\right)\,.
\end{equation}
The saddle point equation $\p_zS(z,m)=0$ leads to the following quadratic equation for the saddle point $z_*$:
\begin{equation}\label{J41_saddle}
	1-\left(m^2-1+m^{-2}\right)z_*+z_*^2=0\,.
\end{equation}
For the longitude coordinate we derive:
\begin{equation}
	\ell=e^{\frac{m}{2}\frac{dS(z_*,m)}{dm}}=\frac{m^2 z_*-1}{m^2-z_*}\,.
\end{equation}
By re-expressing the saddle point value $z_*$ in terms of $\ell$ from this relation and substituting it back into constraint \eqref{J41_saddle}, one arrives at the A-polynomial for the figure-eight knot:
\begin{equation}
	A_{4_1}(\ell,m)=1+\left(-\frac{1}{m^4}+\frac{1}{m^2}+2+m^2-m^4\right)\ell+\ell^2=0\,.
\end{equation}

The figure-eight knot is hyperbolic, so we can calculate its hyperbolic volume in this setting.
If we set $m=1$,\footnote{
The holonomy of the saddle connection around the knot should be a unipotent element in $SL(2,\IC)$ \cite{Garoufalidis_Goerner_Zickert_2015}.
Otherwise, the Thurston hyperbolic metric \cite{Thurston1980} cannot be continued to the cusp containing the knot in question.
This imposes a constraint allowing one to fix $m$:
\begin{equation}
	\Tr\left(\begin{array}{cc}
		1 & * \\
		0 & 1
	\end{array}\right)=2=m+m^{-1}\;\Rightarrow\; m=1\,.
\end{equation} 
} equation \eqref{J41_saddle} has the root $z_*=e^{\frac{\pi\I}{3}}$.
So, the extremal action corresponds at this point to the hyperbolic volume of $4_1$:
\begin{equation}
	\I S(e^{\frac{\pi\I}{3}},1)=\I\left({\rm Li}_2\left(e^{-\frac{\pi\I}{3}}\right)-{\rm Li}_2\left(e^{\frac{\pi\I}{3}}\right)\right)=2.02988\ldots\,.
\end{equation}

The Jones polynomial for $4_1$ satisfies a difference equation \cite{garoufalidis_2004}:
\begin{equation}
\begin{aligned}
&q^{2r}\Big((q^{4r-2}-1)(q^{2r+2}-1)J_{4_1}(r|q) + (q^{4r+2}-1)(q^{2r-2}-1)J_{4_1}(r-2|q)\Big) -\\
&-(q^{4r}-1)(q^{2r}-1)(q^{6r}-q^{4r}-q^{2r+2}-q^{2r-2}-1+q^{-2r})J_{4_1}(r-1|q) = (q^{2r}+1)(q^{4r+2}-1)(q^{4r-2}-1)\,.
\end{aligned}
\end{equation}
By substituting $q^{r}=m$, $J_{\CK}(r+1|q)=\hat\ell J_{\CK}(r|q)$, we will arrive at:
\begin{equation}
	m^2(m^4-1)(m^2-1)\ell^{-2}A_{4_1}(\ell,m)J_{4_1}=(m^2+1)(m^4-1)^2+O(\hbar)\,,
\end{equation}
or
\begin{equation}
	\ell^{-2}A_{4_1}(\ell,m)J_{4_1}=(m+m^{-1})^2+O(\hbar)\,.
\end{equation}
This relation could be further represented in the following form:
\begin{equation}
	\raisebox{-4.5mm}{$\begin{array}{c}
		\begin{tikzpicture}
			\node {$(1-\ell^{-1})\left(\left(m^{-4}-m^{-2}-3-m^2+m^4\right)-\ell^{-1}\right)J_{4_1}+(m+m^{-1})^2\left(\left(-m^{-2}+3-m^2\right)J_{4_1}-1\right)=O(\hbar)\,.$};
			\foreach \x/\y/\h in {-7.9/-0.7/-0.2} {
				\draw[ultra thin, fill] (\x,\h) to[out=270,in=180] (\x+0.3,\h-0.17) -- (0.5*\x+0.5*\y-0.3,\h-0.17) to[out=0,in=90] (0.5*\x+0.5*\y,\h-0.3) to[out=90,in=0] (0.5*\x+0.5*\y-0.3,\h-0.13) -- (\x+0.3,\h-0.13) to[out=180,in=270] (\x,\h);
				\draw[ultra thin, fill] (\y,\h) to[out=270,in=0] (\y-0.3,\h-0.17) -- (0.5*\x+0.5*\y+0.3,\h-0.17) to[out=180,in=90] (0.5*\x+0.5*\y,\h-0.3) to[out=90,in=180] (0.5*\x+0.5*\y+0.3,\h-0.13) -- (\y-0.3,\h-0.13) to[out=0,in=270] (\y,\h);
				\node[below] at (0.5*\x+0.5*\y,\h-0.3) {$\scriptstyle \ell^{-2}\left(A_{4_1}(\ell,m)-A_{4_1}(1,m)\right)$};
			}
			\foreach \x/\y/\h in {0.35/2.15/-0.2} {
				\draw[ultra thin, fill] (\x,\h) to[out=270,in=180] (\x+0.3,\h-0.17) -- (0.5*\x+0.5*\y-0.3,\h-0.17) to[out=0,in=90] (0.5*\x+0.5*\y,\h-0.3) to[out=90,in=0] (0.5*\x+0.5*\y-0.3,\h-0.13) -- (\x+0.3,\h-0.13) to[out=180,in=270] (\x,\h);
				\draw[ultra thin, fill] (\y,\h) to[out=270,in=0] (\y-0.3,\h-0.17) -- (0.5*\x+0.5*\y+0.3,\h-0.17) to[out=180,in=90] (0.5*\x+0.5*\y,\h-0.3) to[out=90,in=180] (0.5*\x+0.5*\y+0.3,\h-0.13) -- (\y-0.3,\h-0.13) to[out=0,in=270] (\y,\h);
				\node[below] at (0.5*\x+0.5*\y,\h-0.3) {$\scriptstyle R_{4_1}(m)$};
			}
			\foreach \x/\y/\h in {2.45/5.15/-0.2} {
				\draw[ultra thin, fill] (\x,\h) to[out=270,in=180] (\x+0.3,\h-0.17) -- (0.5*\x+0.5*\y-0.3,\h-0.17) to[out=0,in=90] (0.5*\x+0.5*\y,\h-0.3) to[out=90,in=0] (0.5*\x+0.5*\y-0.3,\h-0.13) -- (\x+0.3,\h-0.13) to[out=180,in=270] (\x,\h);
				\draw[ultra thin, fill] (\y,\h) to[out=270,in=0] (\y-0.3,\h-0.17) -- (0.5*\x+0.5*\y+0.3,\h-0.17) to[out=180,in=90] (0.5*\x+0.5*\y,\h-0.3) to[out=90,in=180] (0.5*\x+0.5*\y+0.3,\h-0.13) -- (\y-0.3,\h-0.13) to[out=0,in=270] (\y,\h);
				\node[below] at (0.5*\x+0.5*\y,\h-0.3) {$\scriptstyle \Delta_{4_1}(\ell)$};
			}
		\end{tikzpicture}
	\end{array}$}
\end{equation}

We should note that an expansion of the ``action'' near the lower summation limit $z=1$ produces the correct series term in \eqref{J41_naive}:
\begin{equation}
	S(z,\ft)=\log\left(\{\ft\}^2\right)(z-1)+O\left((z-1)^2\right)\,.
\end{equation}

As the final comment for this section, let us note that there are two ways to define the eikonal asymptotics $S(\ft)$.
One way, as we said, is to solve a simple differential equation $\ft \frac{d}{d\ft}S=2\ell(\ft)$ where $\ell$ is a root of the A-polynomial $A(\ell(\ft),\ft)=0$.
The other way is to substitute the solution for the saddle point $z_*$ from \eqref{J41_saddle} directly into \eqref{S41}, which will produce a saddle-point value $S(z_*(\ft),\ft)$.
The compatibility of these two ways leads to the following rather ugly-looking, yet correct integration formula:
\begin{equation}
	\begin{aligned}
	&\int \frac{dt}{t}\,\log\left(\frac{t^8-t^6-2 t^4-t^2+1\pm(t^4-1)\sqrt{t^8-2 t^6-t^4-2 t^2+1}}{2 t^4}\right)=\\
	&={\rm Li}_2\left(\frac{t^4-t^2+1\mp\sqrt{t^8-2 t^6-t^4-2 t^2+1}}{2t^4}\right)-{\rm Li}_2\left(\frac{t^4-t^2+1\pm\sqrt{t^8-2 t^6-t^4-2 t^2+1}}{2t^4}\right)+\\
	&+\log\, t^2\,\log\left(\frac{t^4-t^2+1\pm\sqrt{t^8-2 t^6-t^4-2 t^2+1}}{2 t^2}\right)+C\,.
	\end{aligned}
\end{equation}


\section{Quantum difference equations for knot polynomials}\label{sec:equations}

Quite a useful trick in studying even intrinsically real integrals and differential equations is complexification.
One analytically continues integration variables, integrands, coefficients of differential equations and their solutions to the whole complex plane.
Then, rich methods of saddle point approximations, WKB expansions, etc. in the complex plane are applicable.
This action is much less trivial for QFT path integrals, as reality conditions for the fields entering actions and integration measures have physical meaning.

For 3d Chern-Simons theory with a compact Lie gauge group $G$, complexification could be performed in the following way \cite{Witten:1989ip,Gukov:2003na,Witten:2010cx}.
The action could be rewritten in terms of the Lie algebra $\fg$, which has a natural complexification $\fg_{\IC}$;
therefore, after complexification, we will not be able to study the effects of global topology of $G$ on the path integral, unfortunately.
The action splits into holomorphic and anti-holomorphic parts.
We are interested in the holomorphic part only, assuming the integral is taken along a path in holomorphic field configurations such that the integral converges in a broad sense of convergence for path integrals.
We assume that the integration path is a combination of Lefschetz thimbles; therefore, the path integral reduces to a sum over respective saddles $A_*$, which in Chern-Simons theory are flat connections $F_*=dA_*+A_*\wedge A_*=0$:
\begin{equation}\label{expan}
	\int DA\, e^{\frac{1}{8\pi\hbar'}\int \Tr\left(AdA+\frac{2}{3}A^3\right)}=\sum\lm_{A_*\in {\rm saddles}\,(F_*=0)}c_*\frac{e^{\frac{1}{8\pi\hbar}S_{CS}(A_*)+O(\hbar)}}{\sqrt{{\rm Det\left(\star(d+A_*\wedge)\right)}}}\,,
\end{equation}
where the Chern-Simons level $1/\hbar'$ is allowed to be complex and gets renormalized to $1/\hbar$ in the quantum theory, and coefficients $c_*$ define a decomposition of the integration cycle over Lefschetz thimbles.

\begin{figure}[ht!]
	\centering
	\begin{tikzpicture}
	\node (A) at (0,0) {\scalebox{0.7}{$\begin{array}{c}
			\begin{tikzpicture}
			\begin{scope}[scale=0.6]
			\begin{scope}[shift={(0,1)}]
			\draw[fill=white!50!gray] (0,-1) circle (2.1);
			\begin{scope}[shift={(0,-1)}]
			\begin{scope}[yscale=0.2]
			\draw[thin, dashed] (0,0) circle (2.1);
			\end{scope}
			\end{scope}
			\draw[line width=1mm] (1,0) to[out=90,in=90] (-1,0.15) (-1,-0.15) to[out=270,in=135] (0,-1) to[out=315,in=90] (0.4,-1.5) to[out=270,in=45] (0.1,-1.9) (-0.1,-2.1) to[out=225,in=270] (-1.8,-1.5) to[out=90,in=180] (-1,0) -- (0.85,0) (1.15,0) to[out=0,in=90] (1.8,-1.5) to[out=270, in=315] (0,-2) to[out=135,in=270] (-0.4,-1.5) to[out=90,in=225] (-0.1,-1.1) (0.1,-0.9) to[out=45,in=270] (1,0);
			\draw[line width=0.7mm,white] (1,0) to[out=90,in=90] (-1,0.15) (-1,-0.15) to[out=270,in=135] (0,-1) to[out=315,in=90] (0.4,-1.5) to[out=270,in=45] (0.1,-1.9) (-0.1,-2.1) to[out=225,in=270] (-1.8,-1.5) to[out=90,in=180] (-1,0) -- (0.85,0) (1.15,0) to[out=0,in=90] (1.8,-1.5) to[out=270, in=315] (0,-2) to[out=135,in=270] (-0.4,-1.5) to[out=90,in=225] (-0.1,-1.1) (0.1,-0.9) to[out=45,in=270] (1,0);
			\end{scope}
			\end{scope}
			\end{tikzpicture}
			\end{array}$}};
	\node (B) at (4,0) {\scalebox{0.7}{$\begin{array}{c}
			\begin{tikzpicture}
			\begin{scope}[scale=0.4]
			\draw[thick,fill=white!50!gray] (-2.9,0.) to[out=-89.8399,in=100.017] (-2.87723,-0.26343) to[out=-79.9828,in=118.828] (-2.70107,-0.76265) to[out=-61.1725,in=134.258] (-2.40139,-1.16537) to[out=-45.742,in=146.399] (-2.02811,-1.47359) to[out=-33.6013,in=156.101] (-1.60625,-1.70413) to[out=-23.8987,in=164.195] (-1.14839,-1.86896) to[out=-15.8051,in=171.311] (-0.66453,-1.97387) to[out=-8.68872,in=184.42] (0.33791,-2.01131) to[out=4.41993,in=198.572] (1.30922,-1.81925) to[out=18.5724,in=207.161] (1.75584,-1.63275) to[out=27.1611,in=217.636] (2.16277,-1.37722) to[out=37.6355,in=230.875] (2.51464,-1.03837) to[out=50.8749,in=247.59] (2.7788,-0.60076) to[out=67.5902,in=269.84] (2.9,0.)
			(2.9,0.) to[out=90.1601,in=292.698] (2.7788,0.60076) to[out=112.698,in=309.358] (2.51464,1.03837) to[out=129.358,in=322.547] (2.16277,1.37722) to[out=142.547,in=332.985] (1.75584,1.63275) to[out=152.985,in=341.551] (1.30922,1.81925) to[out=161.551,in=355.683] (0.33791,2.01131) to[out=175.683,in=8.79496] (-0.66453,1.97387) to[out=-171.205,in=15.9226] (-1.14839,1.86896) to[out=-164.077,in=24.036] (-1.60625,1.70413) to[out=-155.964,in=33.7698] (-2.02811,1.47359) to[out=-146.23,in=45.956] (-2.40139,1.16537) to[out=-134.044,in=61.4422] (-2.70107,0.76265) to[out=-118.558,in=80.2966] (-2.87723,0.26343) to[out=-99.7034,in=89.8399] (-2.9,0.);
			\draw[thick,fill=white]
			(-1.11956,0.09908) to[out=-49.9491,in=153.423] (-0.92834,-0.03418) to[out=-26.5774,in=161.919] (-0.72701,-0.11594) to[out=-18.081,in=169.27] (-0.47349,-0.18059) to[out=-10.7297,in=175.99] (-0.18692,-0.21756) to[out=-4.01049,in=184.113] (0.18692,-0.21756) to[out=4.11276,in=190.839] (0.47349,-0.18059) to[out=10.8386,in=198.203] (0.72701,-0.11594) to[out=18.2034,in=206.723] (0.92834,-0.03418) to[out=26.7226,in=229.949] (1.11956,0.09908)
			(0.97323,0.01035) to[out=150.737,in=339.671] (0.78968,0.09405) to[out=159.671,in=347.286] (0.54899,0.16486) to[out=167.286,in=354.139] (0.2697,0.21034) to[out=174.139,in=357.416] (0.12181,0.22137) to[out=177.416,in=1.0626] (-0.04693,0.22375) to[out=-178.937,in=5.96464] (-0.2697,0.21034) to[out=-174.035,in=12.8258] (-0.54899,0.16486) to[out=-167.174,in=20.4567] (-0.78968,0.09405) to[out=-159.543,in=29.2627] (-0.97323,0.01035);
			\draw[thick,\myblue]
			(-2.,0.74437) to[out=-110.513,in=111.519] (-1.99235,0.25547) to[out=-68.4811,in=143.432] (-1.59517,-0.22917) to[out=-36.5678,in=164.608] (-0.84954,-0.57841) to[out=-15.3925,in=173.316] (-0.39091,-0.66766) to[out=-6.68404,in=182.646] (0.15119,-0.68718) to[out=2.64617,in=195.535] (0.84954,-0.57841) to[out=15.5348,in=216.766] (1.59517,-0.22917) to[out=36.7659,in=248.787] (1.99235,0.25547) to[out=68.7869,in=290.513] (2.,0.74437)
			(2.,0.74437) to[out=110.827,in=311.265] (1.85392,0.98337) to[out=131.265,in=335.001] (1.41484,1.29737) to[out=155.001,in=348.618] (0.84207,1.47882) to[out=168.618,in=357.467] (0.21479,1.55296) to[out=177.467,in=5.21746] (-0.42348,1.53888) to[out=-174.783,in=15.2103] (-1.03897,1.4325) to[out=-164.79,in=31.5016] (-1.57881,1.20973) to[out=-148.498,in=43.8516] (-1.79277,1.04724) to[out=-136.148,in=69.173] (-2.,0.74437);
			\node[\myblue, below left] at (-0.84954,-0.57841) {$\scriptstyle \fl$};
			\draw[\myblue, thick]
			(1.27781,-1.82961) to[out=18.1969,in=232.273] (1.36305,-1.76682) to[out=52.2727,in=255.341] (1.42518,-1.63128) to[out=75.3413,in=270.783] (1.4499,-1.39568) to[out=90.783,in=277.359] (1.4373,-1.22373) to[out=97.3594,in=285.102] (1.38531,-0.96416) to[out=105.102,in=295.592] (1.25201,-0.60221) to[out=115.592,in=310.13] (1.05809,-0.29197) to[out=130.13,in=326.502] (0.91736,-0.16139) to[out=146.502,in=17.7616] (0.72219,-0.11751);
			\draw[\myblue, dashed, thick]
			(0.72219,-0.11751) to[out=-161.803,in=69.0245] (0.59176,-0.26297) to[out=-110.975,in=86.938] (0.55164,-0.47351) to[out=-93.062,in=100.849] (0.58046,-0.83433) to[out=-79.151,in=109.283] (0.66132,-1.13397) to[out=-70.717,in=115.835] (0.75151,-1.35224) to[out=-64.1654,in=122.344] (0.84463,-1.5209) to[out=-57.6557,in=135.66] (0.99829,-1.71626) to[out=-44.3405,in=148.487] (1.09512,-1.79371) to[out=-31.5133,in=197.762] (1.27781,-1.82961);
			\node[\myblue, above right] at (1.4499,-1.39568) {$\scriptstyle \fm$};
			\end{scope}
			\end{tikzpicture}
			\end{array}$}};
	\path (A) edge[<->] node[above] {\scriptsize inversion} (B);
	\node(C) at (8,0) {$\begin{array}{c}
		\begin{tikzpicture}[scale=0.7]
		\draw[white,line width = 1.5mm] (-0.5,0) to[out=90,in=270] (1,1.5) to[out=90,in=0] (0,2);
		\draw[gray, line width =  1mm] (-0.5,0) to[out=90,in=270] (1,1.5) to[out=90,in=0] (0,2);
		\begin{scope}[yscale=0.5]
		\draw[thick, \myblue] ([shift={(0:0.3)}]-1.5,0) arc (0:180:0.3);
		\end{scope}
		\node[left, \myblue] at (-1.8,0) {$\fm$};
		\draw[white,line width = 1.5mm] (0,1) to[out=180,in=90] (-1.5,0) to[out=270,in=180] (-0.5,-0.7) to[out=0,in=270] (0.5,0);
		\draw[gray, line width =  1mm] (0,1.2) to[out=180,in=90] (-1.5,0) to[out=270,in=180] (-0.5,-0.7) to[out=0,in=270] (0.5,0);
		\draw[white,line width = 1.5mm] (0,2) to[out=180,in=90] (-1,1.5) to[out=270,in=90] (0.5,0);
		\draw[gray, line width =  1mm] (0,2) to[out=180,in=90] (-1,1.5) to[out=270,in=90] (0.5,0);
		\begin{scope}[xscale=-1]
		\draw[white,line width = 1.5mm] (0,1.2) to[out=180,in=90] (-1.5,0) to[out=270,in=180] (-0.5,-0.7) to[out=0,in=270] (0.5,0);
		\draw[gray, line width =  1mm] (0,1.2) to[out=180,in=90] (-1.5,0) to[out=270,in=180] (-0.5,-0.7) to[out=0,in=270] (0.5,0);
		\end{scope}
		\begin{scope}[yscale=0.5]
		\draw[white, line width = 0.7mm] ([shift={(180:0.3)}]-1.5,0) arc (180:360:0.3);
		\draw[thick, \myblue] ([shift={(180:0.3)}]-1.5,0) arc (180:360:0.3);
		\end{scope}
		\end{tikzpicture}
		\end{array}$};
	\node(D) at (12,0) {$\begin{array}{c}
		\begin{tikzpicture}[scale=0.7]
		\draw[white,line width = 1.5mm] (-0.5,0) to[out=90,in=270] (1,1.5) to[out=90,in=0] (0,2);
		\draw[gray, line width =  1mm] (-0.5,0) to[out=90,in=270] (1,1.5) to[out=90,in=0] (0,2);
		\begin{scope}[shift={(0.2,-0.2)}]
		\draw[thick, white,line width = 0.7mm] (-0.5,0) to[out=90,in=270] (1,1.5) to[out=90,in=0] (0,2);
		\draw[thick, \myblue] (-0.5,0) to[out=90,in=270] (1,1.5) to[out=90,in=0] (0,2);
		\end{scope}
		\draw[white,line width = 1.5mm] (0,1.2) to[out=180,in=90] (-1.5,0) to[out=270,in=180] (-0.5,-0.7) to[out=0,in=270] (0.5,0);
		\draw[gray, line width =  1mm] (0,1.2) to[out=180,in=90] (-1.5,0) to[out=270,in=180] (-0.5,-0.7) to[out=0,in=270] (0.5,0);
		\begin{scope}[shift={(0.2,-0.2)}]
		\draw[thick, white,line width = 0.7mm] (0,1.2) to[out=180,in=90] (-1.5,0) to[out=270,in=180] (-0.5,-0.7) to[out=0,in=270] (0.5,0);
		\draw[thick, \myblue] (0,1.2) to[out=180,in=90] (-1.5,0) to[out=270,in=180] (-0.5,-0.7) to[out=0,in=270] (0.5,0);
		\end{scope}
		\draw[white,line width = 1.5mm] (0,2) to[out=180,in=90] (-1,1.5) to[out=270,in=90] (0.5,0);
		\draw[gray, line width =  1mm] (0,2) to[out=180,in=90] (-1,1.5) to[out=270,in=90] (0.5,0);
		\begin{scope}[shift={(0.2,-0.2)}]
		\draw[thick, white,line width = 0.7mm] (0,2) to[out=180,in=90] (-1,1.5) to[out=270,in=90] (0.5,0);
		\draw[thick, \myblue] (0,2) to[out=180,in=90] (-1,1.5) to[out=270,in=90] (0.5,0);
		\node[right, \myblue] at (1.5,0) {$\fl$};
		\end{scope}
		\begin{scope}[xscale=-1]
		\draw[white,line width = 1.5mm] (0,1.2) to[out=180,in=90] (-1.5,0) to[out=270,in=180] (-0.5,-0.7) to[out=0,in=270] (0.5,0);
		\draw[gray, line width =  1mm] (0,1.2) to[out=180,in=90] (-1.5,0) to[out=270,in=180] (-0.5,-0.7) to[out=0,in=270] (0.5,0);
		\end{scope}
		\begin{scope}[shift={(0.2,-0.2)}]
		\begin{scope}[xscale=-1]
		\draw[thick, white,line width = 0.7mm] (0,1.2) to[out=180,in=90] (-1.5,0) to[out=270,in=180] (-0.5,-0.7) to[out=0,in=270] (0.5,0);
		\draw[thick, \myblue] (0,1.2) to[out=180,in=90] (-1.5,0) to[out=270,in=180] (-0.5,-0.7) to[out=0,in=270] (0.5,0);
		\end{scope}
		\end{scope}
		\end{tikzpicture}
		\end{array}$};
	\end{tikzpicture}
	\caption{Meridian and longitude operators on knot complement $S^3\setminus K$} \label{fig:lambda_mu}
\end{figure}
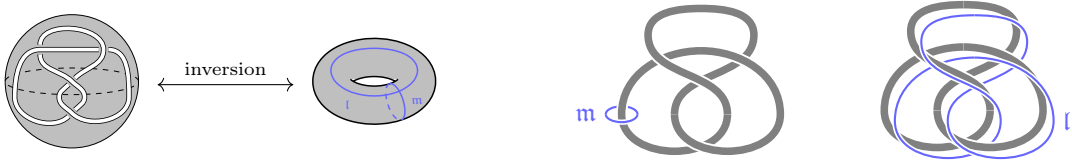

An insertion of knotted Wilson lines does not modify the expansion \eqref{expan} over Lefschetz thimbles.
We expect that an analytic continuation of HOMFLY polynomials for knots also has an asymptotic expansion of the form \eqref{expan} in the quasi-classical limit, where saddles correspond to flat connections in the knot complement with specified boundary holonomies around the knot defect.
In particular, having a flat saddle field configuration $A_*$ satisfying $F_*=0$, one can construct a representation of the knot fundamental group $\pi(\CK)$:
\begin{equation}
	\rho(g):={\rm Pexp}\oint\lm_gA_*\,.
\end{equation}
Let us restrict ourselves to the gauge group $SU(2)$, whose complexification is $SL(2,\IC)$.
One parametrizes the character variety of the fundamental group of the knot complement by fixing eigen values of holonomies for the meridian and the longitude loops (see Fig.~\ref{fig:lambda_mu}):
\begin{equation}
	\Tr\, \fm=m+m^{-1},\quad \Tr\,\fl=\ell+\ell^{-1}\,.
\end{equation}

Canonically, an A-polynomial $A_{\CK}(\ell,m)$ for knot $\CK$ defines \cite{Cooper1994Plane} an algebraic curve in coordinates $(\ell, m)\in \IC^2$ describing possible values of $\ell$ and $m$ for which such a representation $\pi(\CK) \to SL(2,\IC)$ exists.
In other words, it characterizes saddle points in \eqref{expan}.

We would like to mention two specific flat connections in the knot complement.
The first one is called Abelian and corresponds to the case when the representation of the knot fundamental group is concentrated in the Abelian subgroup of $SL(2,\IC)$.
One can easily construct an apparent point in the gauge orbit of this connection for arbitrary $m$ following the Biot-Savart-Laplace law:
\begin{equation}
(A_*)_{\mu}(x)=\log\,m\times\oint\lm_K\frac{\epsilon_{\mu\nu\lambda}dw^{\nu}(x-w)^{\lambda}}{4\pi|x-w|^3}\left(\begin{array}{cc}
1 & 0 \\
0 & -1 \\
\end{array}\right)\,.
\end{equation}
where $x$ is a point in 3d space, and $w$ is a point on the knot.
In this case $S_{CS}(A_*)\equiv 0$.

The other saddle point is canonical for hyperbolic knot complements.
There is a natural decomposition of a 1-form valued in $\fs\fl(2,\IC)$ into real and complex parts:
\begin{equation}
 A_{\mu}=\eta_{ab}e_{\mu}^a\sigma^b+\I \omega_{\mu}^{ab}\epsilon_{abc}\sigma^c\,,
\end{equation}
where $\sigma^a$ are Pauli matrices.
Substituting this ansatz into the flatness condition $F=0$, one obtains the Einstein equation for a hyperbolic metric in terms of the vielbein $e_{\mu}^a$ and a compatibility condition $\nabla e=0$ for the spin connection $\omega_{\mu}^{ab}$.
For hyperbolic knots one can define the hyperbolic Thurston metric \cite{Thurston1980} on the knot complement solving these equations and, therefore, delivering a non-trivial saddle point $A_*$.
In this case, $S_{CS}(A_*)$ produces the hyperbolic volume of the knot $\CK$.

Thinking of the Jones polynomials as analytically continued complex Chern-Simons path integrals on the knot complement allows us to interpret Wilson operators along the meridian cycle and the longitude as the Hopf link and the satellite link, respectively.
This reasoning delivers quantum versions of coordinates $m$ and $\ell$:
\begin{equation}
	\hat m \, J_{\CK}(r|q)=q^r\, J_{\CK}(r|q),\quad \hat \ell \, J_{\CK}(r|q)=J_{\CK}(r+1|q)\,.
\end{equation}
One could obtain difference equations for the Jones polynomials using different methods.
Natural geometric methods produce \emph{homogeneous} difference equations \cite{Galakhov:2024eco,Galakhov:2025ehn,Galakhov:2026xwd}.
We could obtain a homogeneous difference equation from \eqref{main1} in a simple way by dividing both sides by the remnant polynomial and acting with a difference operator $\hat \ell-1$:
\begin{equation}\label{homogen}
	\left(\hat \ell-1\right){\bf A}_{\CK}(\hat \ell,\hat m|q){\bf R}_{\CK}(\hat m|q)^{-1}J_{\CK}(r|q)=0\,.
\end{equation} 

We search for a solution to this equation, expecting the asymptotic behavior of the Jones polynomials to reveal an eikonal form similar to \eqref{expan} for $q=e^{\hbar}$:
\begin{equation}\label{eikon}
	J_{\CK}(r|q)=\frac{e^{\frac{S(m)}{\hbar}+O(\hbar)}}{D(m)},\quad \hat m\,J_{\CK}(r|q)=m\,J_{\CK}(r|q),\quad \hat \ell\,J_{\CK}(r|q)= \ell\,J_{\CK}(r|q)+O(\hbar),\quad \ell=e^{m\frac{dS(m)}{dm}}\,.
\end{equation}
It reduces to an algebraic equation for the A-polynomial with an extra root:
\begin{equation}
	(\ell-1)A_{\CK}(\ell,m)=0\,.
\end{equation}
This extra root $\ell=1$ corresponds to a solution where $S(m)=0$; in other words, it captures the contribution of the \emph{Abelian} connection.
Typically, one omits this root from the canonical definition of the A-polynomial.

We should stress that the \emph{non-homogeneous} equation \eqref{main1} for the Jones polynomials is more informative than the homogeneous one.
As we have already done in Sec.~\ref{sec:toy} in a similar situation, let us substitute ansatz \eqref{eikon} into \eqref{main1}:
\begin{equation}
	A_{\CK}(\ell,m)=R_{\CK}(m)D_{\CK}(m)e^{-\frac{S(m)}{\hbar}}+O(\hbar)\,.
\end{equation}
For solutions with $S(m)\neq 0$, including a calculation of the hyperbolic volume for hyperbolic knots, all the expressions on the r.h.s. are suppressed in the limit $\hbar\to 0$.
The resulting equation allows us to define the behavior of the function $S(m)$ via solving an additional simple differential equation:
\begin{equation}
	m\frac{d}{dm}S(m)=\ell(m)\,,
\end{equation}
where $\ell(m)$ is a root of the A-polynomial $A_{\CK}(\ell,m)=0$.
And for the Abelian connection with $S(m)\equiv 0$, the exponential suppression factor disappears from the r.h.s., and we are able to determine the \emph{next order correction} $D_{\CK}(m)$:
\begin{equation}
	D_{\CK}(m)=\frac{A_{\CK}(1,m)}{R_{\CK}(m)}\,.
\end{equation}
Physically, $D_{\CK}(m)$ for the root $S(m)\equiv 0$ describes a one-loop determinant for fluctuations on the Abelian connection background in expansion \eqref{expan}.
It is known to correspond to the Alexander polynomial of the knot $\CK$.
The reasons why the Alexander polynomial plays the role of the one-loop determinant in the \emph{Abelian} flat connection background and \emph{simultaneously} appears as a factor for $\ell=1$ in the A-polynomial -- which governs the structure of the \emph{non-Abelian} flat connections on the knot complement -- will be discussed in the next section.


\section{Why Alexander?}\label{sec:why}

The fact that $D_{\CK}=\Delta_{\CK}$, where $\Delta_{\CK}$ is the Alexander polynomial of the knot $\CK$, for the branch with $S\equiv 0$ is known as the Melvin-Morton conjecture \cite{Melvin1995}.
Rigorous proofs \cite{Rozansky:1994qe,BarNatan1996,1997CMaPh.183..291R,Garoufalidis:2026mvw} of this observation are rather involved technically.
For example, in \cite{Rozansky:1994qe} L.~Rozansky showed that the one-loop determinant for fluctuations in the Abelian flat connection background is related to the Reidemeister-Ray-Singer torsion of the knot complement; this torsion is known to be proportional to the Alexander polynomial up to a simple factor.
In contrast, the appearance of the Alexander polynomial as a factor in the A-polynomial at $\ell=1$ follows from the fact that roots of $\Delta_{\CK}(\ft)$ ought to be roots of $A_{\CK}(1,\ft)$.
This fact was noted in \cite{Cooper1994Plane} and follows from the Burde-de Rham theorem we will discuss momentarily.

So, to argue that the Melvin-Morton observation is correct we could reverse the logic in the following way.
The fact that the classical limit of the full homogeneous difference equation for the Jones polynomials must have a root $\ell=1$ to take into account the Abelian flat connection might indicate that one could lower the degree of the difference equation by factoring it as \eqref{homogen} and solving the resulting relation as \eqref{main1}.
In this case, as we discussed in Sec.~\ref{sec:equations}, the next order correction in the eikonal approximation to the Jones polynomial asymptotics can be calculated from the A-polynomial at $\ell=1$ and produces the Alexander polynomial.

\begin{figure}[ht!]
	\centering
	\begin{tikzpicture}
		\node(A) at (0,0) {$\begin{array}{c}
				\begin{tikzpicture}
					\begin{scope}[shift={(-1,-0.7)}]
						\begin{scope}[yscale=0.5]
							\draw[thick, \myblue] ([shift={(0:0.35)}]0,0) arc (0:180:0.35);
						\end{scope}
					\end{scope}
					\begin{scope}[shift={(1,-0.7)}]
						\begin{scope}[yscale=0.5]
							\draw[thick, burgundy] ([shift={(0:0.35)}]0,0) arc (0:180:0.35);
						\end{scope}
					\end{scope}
					\draw[black, line width=4.5mm] (1,-1) to[out=90,in=270] (-1,1);
					\draw[white, line width=3.9mm] (1,-1) to[out=90,in=270] (-1,1);
					\begin{scope}[shift={(1,-1)}]
						\begin{scope}[yscale=0.5]
							\draw[draw = none, fill = white] (0,0) circle (0.21);
							\draw[thick, dashed] ([shift={(0:0.21)}]0,0) arc (0:180:0.21);
							\draw[thick] ([shift={(180:0.21)}]0,0) arc (180:360:0.21);
						\end{scope}
					\end{scope}
					\begin{scope}[shift={(-1,1)}]
						\begin{scope}[yscale=0.5]
							\draw[thick, fill = white] (0,0) circle (0.21);
						\end{scope}
					\end{scope}
					\draw[postaction={decorate},decoration={markings, mark= at position 0.9 with {\arrow{stealth}}}] (1,-1) to[out=90,in=270] (-1,1);
					\draw[fill=black] (1,-1) circle (0.04) (-1,1) circle (0.04);
					\draw[black, line width=4.5mm] (-1,-1) to[out=90,in=270] (1,1);
					\draw[white, line width=3.9mm] (-1,-1) to[out=90,in=270] (1,1);
					\begin{scope}[shift={(-1,-1)}]
						\begin{scope}[yscale=0.5]
							\draw[draw = none, fill = white] (0,0) circle (0.21);
							\draw[thick, dashed] ([shift={(0:0.21)}]0,0) arc (0:180:0.21);
							\draw[thick] ([shift={(180:0.21)}]0,0) arc (180:360:0.21);
						\end{scope}
					\end{scope}
					\begin{scope}[shift={(1,1)}]
						\begin{scope}[yscale=0.5]
							\draw[thick, fill = white] (0,0) circle (0.21);
						\end{scope}
					\end{scope}
					\draw[postaction={decorate},decoration={markings, mark= at position 0.9 with {\arrow{stealth}}}] (-1,-1) to[out=90,in=270] (1,1);
					\draw[fill=black] (-1,-1) circle (0.04) (1,1) circle (0.04);
					\draw[thick, \myblue] (0,-1.3) to[out=135,in=270] (-1.35,-0.7);
					\draw[thick, \myblue, postaction={decorate},decoration={markings, mark= at position 0.4 with {\arrowreversed{stealth}}}] (0,-1.3) to[out=120,in=270] (-0.65,-0.7);
					\draw[thick, burgundy] (0,-1.3) to[out=60,in=270] (0.65,-0.7);
					\draw[thick, burgundy, postaction={decorate},decoration={markings, mark= at position 0.4 with {\arrowreversed{stealth}}}] (0,-1.3) to[out=45,in=270] (1.35,-0.7);
					\node[left, \myblue] at (-1.35,-0.7) {$\scriptstyle x_j$};
					\node[right, burgundy] at (1.35,-0.7) {$\scriptstyle x_i$};
				\end{tikzpicture}
			\end{array}$};
		\node(B) at (4,0) {$\begin{array}{c}
				\begin{tikzpicture}
					\begin{scope}[shift={(-1,0.7)}]
						\begin{scope}[yscale=0.5]
							\draw[thick, \mygreen] ([shift={(0:0.35)}]0,0) arc (0:180:0.35);
						\end{scope}
					\end{scope}
					\begin{scope}[shift={(1,0.7)}]
						\begin{scope}[yscale=0.5]
							\draw[thick, \myblue] ([shift={(0:0.35)}]0,0) arc (0:180:0.35);
						\end{scope}
					\end{scope}
					\draw[black, line width=4.5mm] (1,-1) to[out=90,in=270] (-1,1);
					\draw[white, line width=3.9mm] (1,-1) to[out=90,in=270] (-1,1);
					\begin{scope}[shift={(1,-1)}]
						\begin{scope}[yscale=0.5]
							\draw[draw = none, fill = white] (0,0) circle (0.21);
							\draw[thick, dashed] ([shift={(0:0.21)}]0,0) arc (0:180:0.21);
							\draw[thick] ([shift={(180:0.21)}]0,0) arc (180:360:0.21);
						\end{scope}
					\end{scope}
					\begin{scope}[shift={(-1,1)}]
						\begin{scope}[yscale=0.5]
							\draw[thick, fill = white] (0,0) circle (0.21);
						\end{scope}
					\end{scope}
					\draw[postaction={decorate},decoration={markings, mark= at position 0.9 with {\arrow{stealth}}}] (1,-1) to[out=90,in=270] (-1,1);
					\draw[fill=black] (1,-1) circle (0.04) (-1,1) circle (0.04);
					\draw[black, line width=4.5mm] (-1,-1) to[out=90,in=270] (1,1);
					\draw[white, line width=3.9mm] (-1,-1) to[out=90,in=270] (1,1);
					\begin{scope}[shift={(-1,-1)}]
						\begin{scope}[yscale=0.5]
							\draw[draw = none, fill = white] (0,0) circle (0.21);
							\draw[thick, dashed] ([shift={(0:0.21)}]0,0) arc (0:180:0.21);
							\draw[thick] ([shift={(180:0.21)}]0,0) arc (180:360:0.21);
						\end{scope}
					\end{scope}
					\begin{scope}[shift={(1,1)}]
						\begin{scope}[yscale=0.5]
							\draw[thick, fill = white] (0,0) circle (0.21);
						\end{scope}
					\end{scope}
					\draw[postaction={decorate},decoration={markings, mark= at position 0.9 with {\arrow{stealth}}}] (-1,-1) to[out=90,in=270] (1,1);
					\draw[fill=black] (-1,-1) circle (0.04) (1,1) circle (0.04);
					\draw[thick, \mygreen] (0,-1.3) to[out=135,in=270] (-1.35,0.7);
					\draw[thick, \mygreen, postaction={decorate},decoration={markings, mark= at position 0.4 with {\arrowreversed{stealth}}}] (0,-1.3) to[out=120,in=270] (-0.65,0.7);
					\draw[thick, \myblue] (0,-1.3) to[out=60,in=270] (0.65,0.7);
					\draw[thick, \myblue, postaction={decorate},decoration={markings, mark= at position 0.4 with {\arrowreversed{stealth}}}] (0,-1.3) to[out=45,in=270] (1.35,0.7);
					\node[left, \mygreen] at (-1.35,0.7) {$\scriptstyle x_k$};
					\node[right, \myblue] at (1.35,0.7) {$\scriptstyle x_j$};
				\end{tikzpicture}
			\end{array}$};
		\node at (2,0) {$=$};
		\node[below] at  (2,-1.3) {(a)};
		\node(A) at (8,0) {$\begin{array}{c}
				\begin{tikzpicture}
					\begin{scope}[shift={(-1,-0.7)}]
						\begin{scope}[yscale=0.5]
							\draw[thick, burgundy] ([shift={(0:0.35)}]0,0) arc (0:180:0.35);
						\end{scope}
					\end{scope}
					\begin{scope}[shift={(1,-0.7)}]
						\begin{scope}[yscale=0.5]
							\draw[thick, \myblue] ([shift={(0:0.35)}]0,0) arc (0:180:0.35);
						\end{scope}
					\end{scope}
					\draw[black, line width=4.5mm] (-1,-1) to[out=90,in=270] (1,1);
					\draw[white, line width=3.9mm] (-1,-1) to[out=90,in=270] (1,1);
					\begin{scope}[shift={(-1,-1)}]
						\begin{scope}[yscale=0.5]
							\draw[draw = none, fill = white] (0,0) circle (0.21);
							\draw[thick, dashed] ([shift={(0:0.21)}]0,0) arc (0:180:0.21);
							\draw[thick] ([shift={(180:0.21)}]0,0) arc (180:360:0.21);
						\end{scope}
					\end{scope}
					\begin{scope}[shift={(1,1)}]
						\begin{scope}[yscale=0.5]
							\draw[thick, fill = white] (0,0) circle (0.21);
						\end{scope}
					\end{scope}
					\draw[postaction={decorate},decoration={markings, mark= at position 0.9 with {\arrow{stealth}}}] (-1,-1) to[out=90,in=270] (1,1);
					\draw[fill=black] (-1,-1) circle (0.04) (1,1) circle (0.04);
					\draw[black, line width=4.5mm] (1,-1) to[out=90,in=270] (-1,1);
					\draw[white, line width=3.9mm] (1,-1) to[out=90,in=270] (-1,1);
					\begin{scope}[shift={(1,-1)}]
						\begin{scope}[yscale=0.5]
							\draw[draw = none, fill = white] (0,0) circle (0.21);
							\draw[thick, dashed] ([shift={(0:0.21)}]0,0) arc (0:180:0.21);
							\draw[thick] ([shift={(180:0.21)}]0,0) arc (180:360:0.21);
						\end{scope}
					\end{scope}
					\begin{scope}[shift={(-1,1)}]
						\begin{scope}[yscale=0.5]
							\draw[thick, fill = white] (0,0) circle (0.21);
						\end{scope}
					\end{scope}
					\draw[postaction={decorate},decoration={markings, mark= at position 0.9 with {\arrow{stealth}}}] (1,-1) to[out=90,in=270] (-1,1);
					\draw[fill=black] (1,-1) circle (0.04) (-1,1) circle (0.04);
					\draw[thick, burgundy] (0,-1.3) to[out=135,in=270] (-1.35,-0.7);
					\draw[thick, burgundy, postaction={decorate},decoration={markings, mark= at position 0.4 with {\arrowreversed{stealth}}}] (0,-1.3) to[out=120,in=270] (-0.65,-0.7);
					\draw[thick, \myblue] (0,-1.3) to[out=60,in=270] (0.65,-0.7);
					\draw[thick, \myblue, postaction={decorate},decoration={markings, mark= at position 0.4 with {\arrowreversed{stealth}}}] (0,-1.3) to[out=45,in=270] (1.35,-0.7);
					\node[left, burgundy] at (-1.35,-0.7) {$\scriptstyle x_i$};
					\node[right, \myblue] at (1.35,-0.7) {$\scriptstyle x_j$};
				\end{tikzpicture}
			\end{array}$};
		\node(B) at (12,0) {$\begin{array}{c}
				\begin{tikzpicture}
					\begin{scope}[shift={(-1,0.7)}]
						\begin{scope}[yscale=0.5]
							\draw[thick, \myblue] ([shift={(0:0.35)}]0,0) arc (0:180:0.35);
						\end{scope}
					\end{scope}
					\begin{scope}[shift={(1,0.7)}]
						\begin{scope}[yscale=0.5]
							\draw[thick, \mygreen] ([shift={(0:0.35)}]0,0) arc (0:180:0.35);
						\end{scope}
					\end{scope}
					\draw[black, line width=4.5mm] (-1,-1) to[out=90,in=270] (1,1);
					\draw[white, line width=3.9mm] (-1,-1) to[out=90,in=270] (1,1);
					\begin{scope}[shift={(-1,-1)}]
						\begin{scope}[yscale=0.5]
							\draw[draw = none, fill = white] (0,0) circle (0.21);
							\draw[thick, dashed] ([shift={(0:0.21)}]0,0) arc (0:180:0.21);
							\draw[thick] ([shift={(180:0.21)}]0,0) arc (180:360:0.21);
						\end{scope}
					\end{scope}
					\begin{scope}[shift={(1,1)}]
						\begin{scope}[yscale=0.5]
							\draw[thick, fill = white] (0,0) circle (0.21);
						\end{scope}
					\end{scope}
					\draw[postaction={decorate},decoration={markings, mark= at position 0.9 with {\arrow{stealth}}}] (-1,-1) to[out=90,in=270] (1,1);
					\draw[fill=black] (-1,-1) circle (0.04) (1,1) circle (0.04);
					\draw[black, line width=4.5mm] (1,-1) to[out=90,in=270] (-1,1);
					\draw[white, line width=3.9mm] (1,-1) to[out=90,in=270] (-1,1);
					\begin{scope}[shift={(1,-1)}]
						\begin{scope}[yscale=0.5]
							\draw[draw = none, fill = white] (0,0) circle (0.21);
							\draw[thick, dashed] ([shift={(0:0.21)}]0,0) arc (0:180:0.21);
							\draw[thick] ([shift={(180:0.21)}]0,0) arc (180:360:0.21);
						\end{scope}
					\end{scope}
					\begin{scope}[shift={(-1,1)}]
						\begin{scope}[yscale=0.5]
							\draw[thick, fill = white] (0,0) circle (0.21);
						\end{scope}
					\end{scope}
					\draw[postaction={decorate},decoration={markings, mark= at position 0.9 with {\arrow{stealth}}}] (1,-1) to[out=90,in=270] (-1,1);
					\draw[fill=black] (1,-1) circle (0.04) (-1,1) circle (0.04);
					\draw[thick, \myblue] (0,-1.3) to[out=135,in=270] (-1.35,0.7);
					\draw[thick, \myblue, postaction={decorate},decoration={markings, mark= at position 0.4 with {\arrowreversed{stealth}}}] (0,-1.3) to[out=120,in=270] (-0.65,0.7);
					\draw[thick, \mygreen] (0,-1.3) to[out=60,in=270] (0.65,0.7);
					\draw[thick,\mygreen, postaction={decorate},decoration={markings, mark= at position 0.4 with {\arrowreversed{stealth}}}] (0,-1.3) to[out=45,in=270] (1.35,0.7);
					\node[left, \myblue] at (-1.35,0.7) {$\scriptstyle x_j$};
					\node[right, \mygreen] at (1.35,0.7) {$\scriptstyle x_k$};
				\end{tikzpicture}
			\end{array}$};
		\node at (10,0) {$=$};
		\node[below] at  (10,-1.3) {(b)};
	\end{tikzpicture}
	\caption{Relations for the Wirtinger basis: (a) $x_ix_j=x_jx_k$, (b) $x_jx_i=x_kx_j$.}\label{fig:Wirtinger}
\end{figure}
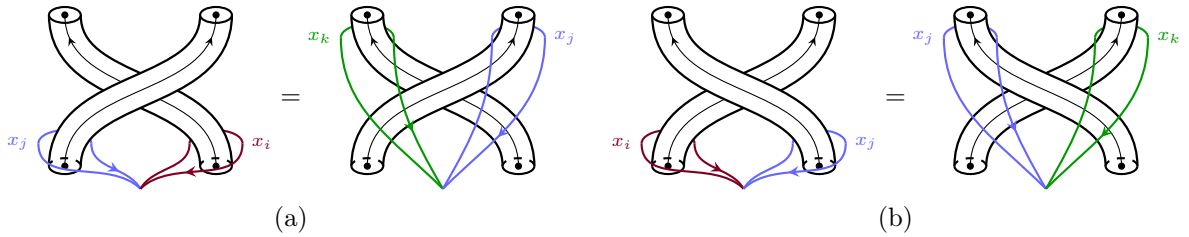

In the rest of this section, we review the Burde-de Rham theorem following \cite{may_2006,silver_williams_2011}.
One should start with the knot group $\pi(\CK)$ constructed as a fundamental group of the knot complement $\pi_1(S^3\setminus \CK)$.
It is easy to give a Wirtinger presentation for this group.
One assumes that the basepoint is at the observer's eye, looking at the knot diagram drawn on a piece of paper.
Then one chooses generators $x_i$ of $\pi(\CK)$ as loops encircling all the arcs of the knot diagram (as seen from observer's viewpoint), each loop forming either the right-hand or the left-hand grip with the knot strand.
Here, without loss of generality, we choose the left one.
Then at the crossings one imposes obvious relations $r_i$ for generators following from loop homotopy (see Fig.~\ref{fig:Wirtinger}), where one of the relations is always redundant, so that
\begin{equation}
	\pi(K)=\left\langle x_1,x_2,\ldots, x_{\#\;{\rm of}\;{\rm arcs}}\big| r_1,r_2,\ldots,r_{\#\;{\rm of}\;{\rm ints}-1}  \right\rangle\,.
\end{equation} 

Classically, A-polynomials study character varieties for $\pi(\CK)$, in particular homomorphisms $\pi(\CK)\to SL(2,\IC)$ modulo  $SL(2,\IC)$.
Following de Rham \cite{de_rham_1967} we choose a \emph{subgroup} $\Gamma\subset SL(2,\IC)$ of \emph{upper-triangular} matrices.
All the loops $x_i$ represent meridian holonomies, therefore:
\begin{equation}
	\Tr\,x_i=m+m^{-1}\,.
\end{equation}
Moreover, relations depicted in Fig.~\ref{fig:Wirtinger} impose the following possible form for a representation of $x_i$ in $\Gamma$:
\begin{equation}
	\rho(x_i)=\left(\begin{array}{cc}
		m & u_i\\
		0 & m^{-1}\\
	\end{array}\right)\,,
\end{equation}
for some moduli $u_i$.
Relations $r_i$ lead to linear relations among moduli $u_i$:
\begin{equation}
	\mbox{Fig.~\ref{fig:Wirtinger}(a): }\frac{u_i}{m}+mu_j=\frac{u_j}{m}+u_k,\quad\quad  \mbox{Fig.~\ref{fig:Wirtinger}(b): }\frac{u_j}{m}+mu_i=\frac{u_k}{m}+m u_j\,.
\end{equation}
A constraint for this system to have a non-Abelian solution, when all the matrices $\rho(x_i)$ cannot be diagonalized simultaneously,  is given by the Alexander polynomial of the knot $\CK$:
\begin{equation}
	\Delta_{\CK}(m)=0\,.
\end{equation}

For example, consider the figure-eight knot with the following parametrization of Wirtinger elements:
\begin{equation}
	\begin{array}{c}
		\begin{tikzpicture}
			\tikzset{
				col1/.style={white!60!palette2},
				col2/.style={white!60!palette3},
				col3/.style={white!60!palette4},
				col4/.style={white!60!palette7},
				col5/.style={palette2},
				col6/.style={palette3},
				col7/.style={palette4},
				col8/.style={palette7},
				col9/.style={black},
			}
			\path[
			decoration={
				markings,
				mark=at position 1 with {
					\xdef\totallength{\pgfdecoratedpathlength}
				}
			},
			decorate] (0,1) to[out=180,in=90] (-1.5,-0.3) to[out=270,in=180] (-0.8,-1);
			\pgfmathsetmacro\firstpart{0.3*\totallength}
			\pgfmathsetmacro\secondpart{0.7*\totallength}
			\draw[line width = 2.9mm] (0,1) to[out=180,in=90] (-1.5,-0.3) to[out=270,in=180] (-0.8,-1);
			\draw[
			col1,
			dash pattern=on \firstpart pt off 1000pt,
			dash phase=0pt,
			line width = 2.5mm
			] (0,1) to[out=180,in=90] (-1.5,-0.3) to[out=270,in=180] (-0.8,-1);
			\draw[
			col2,
			dash pattern=on \secondpart pt off 1000pt,
			dash phase=-\firstpart pt,
			line width = 2.5mm
			](0,1) to[out=180,in=90] (-1.5,-0.3) to[out=270,in=180] (-0.8,-1);
			\draw[thin, col9, postaction={decorate},decoration={markings, mark= at position 0.9 with {\arrow{stealth}}}] (0,1) to[out=180,in=90] (-1.5,-0.3) to[out=270,in=180] (-0.8,-1);
			\path[
			decoration={
				markings,
				mark=at position 1 with {
					\xdef\totallength{\pgfdecoratedpathlength}
				}
			},
			decorate] (0.6,-0.3) to[out=90,in=270] (-1,1) to[out=90,in=180] (0,1.6) to[out=0,in=90] (1,1);
			\pgfmathsetmacro\firstpart{0.18*\totallength}
			\pgfmathsetmacro\secondpart{0.82*\totallength}
			\draw[line width = 2.9mm] (0.6,-0.3) to[out=90,in=270] (-1,1) to[out=90,in=180] (0,1.6) to[out=0,in=90] (1,1);
			\draw[
			col2,
			dash pattern=on \firstpart pt off 1000pt,
			dash phase=0pt,
			line width = 2.5mm
			]  (0.6,-0.3) to[out=90,in=270] (-1,1) to[out=90,in=180] (0,1.6) to[out=0,in=90] (1,1);
			\draw[
			col3,
			dash pattern=on \secondpart pt off 1000pt,
			dash phase=-\firstpart pt,
			line width = 2.5mm
			]  (0.6,-0.3) to[out=90,in=270] (-1,1) to[out=90,in=180] (0,1.6) to[out=0,in=90] (1,1);
			\draw[thin, col9, postaction={decorate},decoration={markings, mark= at position 0.9 with {\arrow{stealth}}}] (0.6,-0.3) to[out=90,in=270] (-1,1) to[out=90,in=180] (0,1.6) to[out=0,in=90] (1,1);
			\path[
			decoration={
				markings,
				mark=at position 1 with {
					\xdef\totallength{\pgfdecoratedpathlength}
				}
			},
			decorate]  (1,1) to[out=270,in=90] (-0.6,-0.3);
			\pgfmathsetmacro\firstpart{0.07*\totallength}
			\pgfmathsetmacro\secondpart{0.93*\totallength}
			\draw[line width = 2.9mm]    (1,1.01) to[out=270,in=90] (-0.6,-0.3);
			\draw[
			col3,
			dash pattern=on \firstpart pt off 1000pt,
			dash phase=0pt,
			line width = 2.5mm
			]   (1,1.01) to[out=270,in=90] (-0.6,-0.3);
			\draw[
			col4,
			dash pattern=on \secondpart pt off 1000pt,
			dash phase=-\firstpart pt,
			line width = 2.5mm
			]  (1,1.01) to[out=270,in=90] (-0.6,-0.3);
			\draw[col9, thin, postaction={decorate},decoration={markings, mark= at position 0.9 with {\arrow{stealth}}}]  (1,1.01) to[out=270,in=90] (-0.6,-0.3);
			\path[
			decoration={
				markings,
				mark=at position 1 with {
					\xdef\totallength{\pgfdecoratedpathlength}
				}
			},
			decorate] (-0.6,-0.3) to[out=270,in=180] (0.8,-1) to[out=0,in=270] (1.5,-0.3);
			\pgfmathsetmacro\firstpart{0.3*\totallength}
			\pgfmathsetmacro\secondpart{0.7*\totallength}
			\draw[line width = 2.9mm]   (-0.6,-0.3) to[out=270,in=180] (0.8,-1) to[out=0,in=270] (1.5,-0.3);
			\draw[
			col4,
			dash pattern=on \firstpart pt off 1000pt,
			dash phase=0pt,
			line width = 2.5mm
			] (-0.6,-0.29) to[out=270,in=180] (0.8,-1) to[out=0,in=270] (1.5,-0.3);
			\draw[
			col1,
			dash pattern=on \secondpart pt off 1000pt,
			dash phase=-\firstpart pt,
			line width = 2.5mm
			] (-0.6,-0.29) to[out=270,in=180] (0.8,-1) to[out=0,in=270] (1.5,-0.3);
			\draw[thin, col9]  (-0.6,-0.29) to[out=270,in=180] (0.8,-1) to[out=0,in=270] (1.5,-0.3);
			\draw[line width = 2.9mm]   (1.5,-0.3) to[out=90,in=0] (0,1);
			\draw[line width = 2.5mm, col1] (1.5,-0.31) to[out=90,in=0] (0,1);
			\draw[thin, col9, postaction={decorate},decoration={markings, mark= at position 0.3 with {\arrow{stealth}}}]   (1.5,-0.31) to[out=90,in=0] (0,1);
			\draw[line width = 2.9mm]   (-0.8,-1) to[out=0,in=270] (0.6,-0.3);
			\draw[line width = 2.5mm,col2] (-0.8,-1) to[out=0,in=270] (0.6,-0.29);
			\draw[thin, col9] (-0.8,-1) to[out=0,in=270] (0.6,-0.29);
			\begin{scope}[shift={(-1.5,-0.3)}]
				\begin{scope}[yscale = 0.5, scale=0.7]
					\draw[col9, dashed] ([shift={(0:0.21)}]0,0) arc (0:180:0.21);
					\draw[col9] ([shift={(180:0.21)}]0,0) arc (180:360:0.21);
				\end{scope}
			\end{scope}
			\begin{scope}[shift={(1.5,-0.3)}]
				\begin{scope}[yscale = 0.5, scale=0.7]
					\draw[col9, dashed] ([shift={(0:0.21)}]0,0) arc (0:180:0.21);
					\draw[col9] ([shift={(180:0.21)}]0,0) arc (180:360:0.21);
				\end{scope}
			\end{scope}
			\begin{scope}[shift={(-0.6,-0.3)}]
				\begin{scope}[yscale = 0.5, scale=0.7]
					\draw[col9, dashed] ([shift={(0:0.21)}]0,0) arc (0:180:0.21);
					\draw[col9] ([shift={(180:0.21)}]0,0) arc (180:360:0.21);
				\end{scope}
			\end{scope}
			\begin{scope}[shift={(0.6,-0.3)}]
				\begin{scope}[yscale = 0.5, scale=0.7]
					\draw[col9, dashed] ([shift={(0:0.21)}]0,0) arc (0:180:0.21);
					\draw[col9] ([shift={(180:0.21)}]0,0) arc (180:360:0.21);
				\end{scope}
			\end{scope}
			\begin{scope}[shift={(-1,1)}]
				\begin{scope}[yscale = 0.5, scale=0.7]
					\draw[col9, dashed] ([shift={(0:0.21)}]0,0) arc (0:180:0.21);
					\draw[col9] ([shift={(180:0.21)}]0,0) arc (180:360:0.21);
				\end{scope}
			\end{scope}
			\begin{scope}[shift={(1,1)}]
				\begin{scope}[yscale = 0.5, scale=0.7]
					\draw[col9, dashed] ([shift={(0:0.21)}]0,0) arc (0:180:0.21);
					\draw[col9] ([shift={(180:0.21)}]0,0) arc (180:360:0.21);
				\end{scope}
			\end{scope}
			\begin{scope}[shift={(0,1.6)}]
				\begin{scope}[rotate=-90]
					\begin{scope}[yscale = 0.5, scale=0.7]
						\draw[col9, dashed] ([shift={(0:0.21)}]0,0) arc (0:180:0.21);
						\draw[col9] ([shift={(180:0.21)}]0,0) arc (180:360:0.21);
					\end{scope}
				\end{scope}
			\end{scope}
			\begin{scope}[shift={(0,1)}]
				\begin{scope}[rotate=-90]
					\begin{scope}[yscale = 0.5, scale=0.7]
						\draw[col9, dashed] ([shift={(0:0.21)}]0,0) arc (0:180:0.21);
						\draw[col9] ([shift={(180:0.21)}]0,0) arc (180:360:0.21);
					\end{scope}
				\end{scope}
			\end{scope}
			\node[left, col6] at (-1.6,-0.3) {$\scriptstyle x_1$};
			\node[right, col5] at (1.6,-0.3) {$\scriptstyle x_4$};
			\node[left, col8] at (-0.7,-0.3) {$\scriptstyle x_3$};
			\node[left, col7] at (-1.1,1) {$\scriptstyle x_2$};
		\end{tikzpicture}
	\end{array}\quad {\renewcommand{\arraystretch}{1.5}
		\begin{array}{c}
			r_1:\quad m u_1-m u_3+\dfrac{u_3}{m}-\dfrac{u_2}{m}=0\,,\\
			r_2:\quad -m u_1+\dfrac{u_1}{m}+m u_3-\dfrac{u_4}{m}=0\,,\\
			r_3:\quad -m u_1+m u_2+\dfrac{u_4}{m}-\dfrac{u_2}{m}=0\,,\\
			r_4:\quad \dfrac{u_2}{m}-m u_3+m u_4-\dfrac{u_4}{m}=0\,.
		\end{array}
	}
\end{equation}
Apparently, the determinant of the matrix describing this system is 0.
This implies that a solution has an affine gauge symmetry $u_i\to u_i+v_i$, where $v_i$ is a null vector, and, as we said earlier, the system of equations $r_1,\ldots,r_4$ for group elements of $\pi(\CK)$ is redundant.
We use this symmetry to set $u_4=0$ and drop $r_4$.
The remaining system has a solution if 
\begin{equation}
	m\,\Delta_{4_1}(m)={\rm Det}\left(
	\begin{array}{ccc}
		m & -m^{-1} & m^{-1}-m \\
		m^{-1}-m & 0 & m \\
		-m & m-m^{-1} & 0 \\
	\end{array}
	\right)=\left(3-m^2-\frac{1}{m^2}\right)m=0\,.
\end{equation}

The fact that a constraint for a system $r_i=0$ to have a non-trivial solution in this case is determined by the knot Alexander polynomial is easy to prove.
It is sufficient to show that the respective determinant satisfies Alexander skein relations.
To do so, let us consider knot diagrams differing in a single intersection and the respective constraints.
We add extra generators and relations to the intersections, to make equations more uniform:
\begin{equation}
	\begin{aligned}
		&\begin{array}{c}
			\begin{tikzpicture}[scale=0.6]
				\draw[thick, -stealth] (0.5,-0.5) -- (-0.5,0.5);
				\draw[white, line width = 2mm] (-0.5,-0.5) -- (0.5,0.5);
				\draw[thick, -stealth] (-0.5,-0.5) -- (0.5,0.5);
				\node[left] at (-0.5,-0.5) {$\scriptstyle x_1$};
				\node[right] at (0.5,-0.5) {$\scriptstyle x_2$};
				\node[left] at (-0.5,0.5) {$\scriptstyle x_3$};
				\node[right] at (0.5,0.5) {$\scriptstyle x_4$};
			\end{tikzpicture}
		\end{array},\quad \begin{array}{c}
			x_2x_1=x_4x_3\,,\\
			x_1=x_4\,,
		\end{array}\quad 
		\Delta\hspace{-1mm}_{\scalebox{0.4}{$\begin{array}{c}
					\begin{tikzpicture}[scale=0.6]
						\draw[ultra thick, -stealth] (0.5,-0.5) -- (-0.5,0.5);
						\draw[white, line width = 2mm] (-0.5,-0.5) -- (0.5,0.5);
						\draw[ultra thick, -stealth] (-0.5,-0.5) -- (0.5,0.5);
					\end{tikzpicture}
				\end{array}$}}\hspace{-1mm}(m)={\rm Det}\left(\begin{array}{ccccccc}
			m & m^{-1} & -m & -m^{-1} & 0 &\ldots & 0\\
			1 & 0 & 0 & -1 & 0 &\ldots & 0\\
			\ldots & \ldots &\ldots &\ldots &\ldots &\ldots &\ldots
		\end{array}\right)=\\
		&\hspace{7cm}=m M_{12}+\frac{M_{13}}{m}+\left(\frac{1}{m}-m\right) M_{23}-m M_{24}-\frac{M_{34}}{m}\,,\\
		&\begin{array}{c}
			\begin{tikzpicture}[scale=0.6]
				\draw[thick, -stealth] (-0.5,-0.5) -- (0.5,0.5);
				\draw[white, line width = 2mm] (0.5,-0.5) -- (-0.5,0.5);
				\draw[thick, -stealth] (0.5,-0.5) -- (-0.5,0.5);				
				\node[left] at (-0.5,-0.5) {$\scriptstyle x_1$};
				\node[right] at (0.5,-0.5) {$\scriptstyle x_2$};
				\node[left] at (-0.5,0.5) {$\scriptstyle x_3$};
				\node[right] at (0.5,0.5) {$\scriptstyle x_4$};
			\end{tikzpicture}
		\end{array},\quad \begin{array}{c}
			x_2x_1=x_4x_3\,,\\
			x_2=x_3\,,
		\end{array}\quad 
		\Delta\hspace{-1mm}_{\scalebox{0.4}{$\begin{array}{c}
					\begin{tikzpicture}[scale=0.6]
						\draw[ultra thick, -stealth] (-0.5,-0.5) -- (0.5,0.5);
						\draw[white, line width = 2mm] (0.5,-0.5) -- (-0.5,0.5);
						\draw[ultra thick, -stealth] (0.5,-0.5) -- (-0.5,0.5);	
					\end{tikzpicture}
				\end{array}$}}\hspace{-1mm}(m)={\rm Det}\left(\begin{array}{ccccccc}
			m & m^{-1} & -m & -m^{-1} & 0 &\ldots & 0\\
			0 & 1 & 1 & 0 & 0 &\ldots & 0\\
			\ldots & \ldots &\ldots &\ldots &\ldots &\ldots &\ldots
		\end{array}\right)=\\
		&\hspace{7cm}=\frac{M_{12}}{m}+\frac{M_{13}}{m}+\left(\frac{1}{m}-m\right) M_{14}-m M_{24}-m M_{34}\,,\\
		&\begin{array}{c}
			\begin{tikzpicture}[scale=0.6]
				\draw[thick, -stealth] (-0.5,-0.5) to[out=45,in=315] (-0.5,0.5);
				\draw[thick, -stealth] (0.5,-0.5) to[out=135,in=225] (0.5,0.5);	
				\node[left] at (-0.5,-0.5) {$\scriptstyle x_1$};
				\node[right] at (0.5,-0.5) {$\scriptstyle x_2$};
				\node[left] at (-0.5,0.5) {$\scriptstyle x_3$};
				\node[right] at (0.5,0.5) {$\scriptstyle x_4$};
			\end{tikzpicture}
		\end{array},\quad \begin{array}{c}
			x_2x_1=x_4x_3\,,\\
			x_1=x_4\,,
		\end{array}\quad 
		\Delta\hspace{-1mm}_{\scalebox{0.4}{$\begin{array}{c}
					\begin{tikzpicture}[scale=0.6]
						\draw[ultra thick, -stealth] (-0.5,-0.5) to[out=45,in=315] (-0.5,0.5);
						\draw[ultra thick, -stealth] (0.5,-0.5) to[out=135,in=225] (0.5,0.5);	
					\end{tikzpicture}
				\end{array}$}}\hspace{-1mm}(m)={\rm Det}\left(\begin{array}{ccccccc}
			1 & 0 & -1 & 0 & 0 &\ldots & 0\\
			0 & 1 & 0 & -1 & 0 &\ldots & 0\\
			\ldots & \ldots &\ldots &\ldots &\ldots &\ldots &\ldots
		\end{array}\right)\hspace{-1mm}=\\
		&\hspace{7cm}=M_{12}+M_{14}-M_{23}+M_{34}\,,
	\end{aligned}
\end{equation}
where $M_{ij}$ are minors with two rows and the $i^{\rm th}$ and the $j^{\rm th}$ columns erased.
Since the rest of diagrams except for this intersection are identical, these minors are equal in all three cases.
Then we arrive at the Alexander skein relation:
\begin{equation}
	\Delta\hspace{-1mm}_{\scalebox{0.4}{$\begin{array}{c}
				\begin{tikzpicture}[scale=0.6]
					\draw[ultra thick, -stealth] (0.5,-0.5) -- (-0.5,0.5);
					\draw[white, line width = 2mm] (-0.5,-0.5) -- (0.5,0.5);
					\draw[ultra thick, -stealth] (-0.5,-0.5) -- (0.5,0.5);
				\end{tikzpicture}
			\end{array}$}}\hspace{-1mm}(m)-\Delta\hspace{-1mm}_{\scalebox{0.4}{$\begin{array}{c}
				\begin{tikzpicture}[scale=0.6]
					\draw[ultra thick, -stealth] (-0.5,-0.5) -- (0.5,0.5);
					\draw[white, line width = 2mm] (0.5,-0.5) -- (-0.5,0.5);
					\draw[ultra thick, -stealth] (0.5,-0.5) -- (-0.5,0.5);	
				\end{tikzpicture}
			\end{array}$}}\hspace{-1mm}(m)=\left(m^{-1}-m\right)\;\Delta\hspace{-1mm}_{\scalebox{0.4}{$\begin{array}{c}
				\begin{tikzpicture}[scale=0.6]
					\draw[ultra thick, -stealth] (-0.5,-0.5) to[out=45,in=315] (-0.5,0.5);
					\draw[ultra thick, -stealth] (0.5,-0.5) to[out=135,in=225] (0.5,0.5);	
				\end{tikzpicture}
			\end{array}$}}\hspace{-1mm}(m)\,.
\end{equation}

Now we turn to the longitude. 
In the case of $4_1$, the longitude in this setting reads:
\begin{equation}
	L=x_2^{-1}x_1x_4^{-1}x_3,\quad \rho(L)=\left(
	\begin{array}{cc}
		1 & \frac{u_1-u_2+u_3-u_4}{m} \\
		0 & 1 \\
	\end{array}
	\right)=:\left(
	\begin{array}{cc}
		\ell & *\\
		0 & \ell^{-1} \\
	\end{array}
	\right)\,.
\end{equation}
We conclude that this representation of holonomies $x_i$, if it exists, corresponds to $\ell=1$.\footnote{A proof that for a generic knot this representation of holonomies corresponds to $\ell=1$ relies on the topological fact that the longitude is an element of the second derived subgroup of the knot group \cite{Cooper1994Plane}.}

Let us recall that the vanishing set of a knot A-polynomial $A_{\CK}(\ell,m)$ contains values of $\ell$ and $m$ such that there are corresponding Chern-Simons saddle-point \emph{flat} connections $A_*$, delivering the homomorphism $\pi(\CK)\to SL(2,\IC)$ via:
\begin{equation}
	x_i={\rm Pexp}\oint\lm_{\gamma_i}A_*\,,
\end{equation}
with constraints on the meridian and the longitude holonomies:
\begin{equation}
	\Tr \,x_A=m+m^{-1},\quad \Tr\,x_B=\ell+\ell^{-1}\,,
\end{equation}
and we have just constructed such a representation for $\ell=1$ and $m$ being a root of the Alexander polynomial $\Delta_{\CK}(m)$.
This implies that if $m_*$ is a root of $\Delta_{\CK}(m_*)=0$, it must be a root of $A_{\CK}(\ell=1,m_*)=0$.
This, in turn, implies that the A-polynomial at value $\ell=1$ is expected to factorize:
\begin{equation}
	A_{\CK}(1,m)=\Delta_{\CK}(m)\, R_{\CK}(m)\,,
\end{equation}
for some remnant polynomial $R_{\CK}$.
Combined with the Melvin-Morton observation, this relation leads to \eqref{main2}.


\section{Conclusion}

We have explained at a rather concrete level the ambiguity of quasi-classical calculations, associated with different saddle points and different solutions of the Ward identities.
Our main interest was the case where the action for one of the branches vanishes and the answer looks like a pure perturbative series in the inverse determinant.
This is exactly the case that occurs in the study of the Kashaev limit of knot polynomials, where the determinant is actually expressed through the Alexander polynomial.
Another, more standard branch in this case is expanded around the exponent of the classical action, which in the case of Jones polynomials is actually the hyperbolic volume – this is the original version of the Kashaev limit.

The central claim is \eqref{main1} and \eqref{main2}, which establish an exact relation between the non-homogeneous version of the quantum ``$A$-polynomial'' (equation for Jones polynomials with respect to representation size) in the quasi-classical limit and the Alexander polynomial.
We presented a detailed consideration for the figure-eight knot, but the claim is that \eqref{main1}, \eqref{main2} are true for all knots.

Although different pieces of this story are well known to the experts and reflected in the existing literature \cite{Rozansky:1994qe,BarNatan1996,1997CMaPh.183..291R,Garoufalidis:2026mvw}, we think that our concise pedagogical summary can be useful and suggests new studies, including other knots and generalizations from Jones to HOMFLY, Kauffman and other knot polynomials, along the lines of \cite{KLM}.
Especially interesting would be a reformulation in terms of
$C$-polynomials \cite{GaroufalidisLin2006,Mironov:2020ecl}, which are simpler and, in a sense, more fundamental – but the way the transition from 
$A$ to $C$ eliminates the knowledge about the Alexander polynomial remains a puzzle.


\section*{Acknowledgments}

The work was supported by the state assignment of the Institute for Information Transmission Problems of RAS.


\bibliographystyle{utphys}
\bibliography{biblio}

\end{document}